\documentclass[journal]{IEEEtran}
\usepackage{url}            
\usepackage{booktabs}       
\usepackage{amsfonts}       
\usepackage{nicefrac}       
\usepackage{microtype}      
\usepackage{lipsum}
\usepackage{amsmath}
\usepackage{graphicx}
\usepackage{subcaption}
\usepackage{graphicx}
\usepackage{amssymb}
\usepackage{pdfpages}
\usepackage{float}
\usepackage{subfiles} 
\usepackage{amsthm}
\usepackage{mathtools}
\usepackage{mathrsfs}

\newtheorem{remark}{Remark}

\DeclareUnicodeCharacter{221E}{-}

\usepackage{algorithm}
\usepackage{algpseudocode}
\usepackage{marginnote}
\usepackage[colorinlistoftodos]{todonotes}
\newcommand\ignore[1]{{}}
\algrenewcommand\algorithmicrequire{\textbf{Input:}}
\algrenewcommand\algorithmicensure{\textbf{Output:}}
\newcommand{\dx}[1]{\textcolor{red}{#1}}

\usepackage{derivative}
\DeclareMathOperator{\st}{s.t.}
\usepackage{multirow}
\usepackage{siunitx}
\usepackage{physics}
\usepackage{cite}

\title{\LARGE \bf
Safe Feedback Motion Planning in Unknown Environments: \\
An Instantaneous Local Control Barrier Function Approach
}

\author{Cong Li,
        Zengjie Zhang{$^*$},
        Nesrin Ahmed,
        Qingchen Liu,
        Fangzhou Liu,
        and~Martin Buss
\thanks{\dx{This work has been accepted for publication in Journal of Intelligent and Robotic Systems}}
\thanks{* Corresponding author}
\thanks{C.Li, N.Ahmed, and M.Buss are with the Chair of Automatic Control Engineering, Technical University of Munich, Theresienstr. 90,80333, Munich, Germany
(e-mail: cong.lea@hotmail.com, \{n.ahmed, mb\}@tum.de).
}
\thanks{Z. Zhang is with the Department of Electrical Engineering, Eindhoven University of Technology, 5600 MB Eindhoven, The Netherlands, e-mail: z.zhang3@tue.nl.}
\thanks{Q. Liu is with the Department of Automation, University of Science and Technology of China, 230027,  Hefei, China, e-mail: qingchen\_liu@ustc.edu.cn.}
\thanks{F. Liu is with the Research Institute of Intelligent Control and Systems, Harbin Institute of Technology, 150001,  Harbin, China, e-mail: fangzhou.liu@hit.edu.cn.}
}

\begin{document}
\maketitle
\begin{abstract}
Mobile robots are desired with resilience to
safely interact with prior-unknown environments and finally accomplish given tasks.
This paper utilizes instantaneous local sensory data to stimulate the safe feedback motion planning (SFMP) strategy with adaptability to diverse prior-unknown environments without building a global map.
This is achieved by the numerical optimization with the constraints, referred to as instantaneous local control barrier functions (IL-CBFs) and goal-driven control Lyapunov functions (GD-CLFs), learned from perceptional signals.
In particular, the IL-CBFs reflecting potential collisions and GD-CLFs encoding incrementally discovered subgoals are first online learned from local perceptual data.
Then, the learned IL-CBFs are united with GD-CLFs in the context of quadratic programming (QP) to generate the safe feedback motion planning strategy.
Rather importantly, an optimization over the admissible control space of IL-CBFs is conducted to enhance the solution feasibility of QP.
The SFMP strategy is developed with theoretically guaranteed collision avoidance and convergence to destinations.
Numerical simulations are conducted to reveal the effectiveness of the proposed SFMP strategy that drives mobile robots to safely reach the destination incrementally in diverse prior-unknown environments.
\end{abstract}

\begin{IEEEkeywords}
Safe feedback motion planning, collision avoidance, instantaneous local control barrier function, goal-driven control Lyapunov function. 
\end{IEEEkeywords}

\section{Introduction}
The safe operation of mobile robots in prior-unknown environments is important in applications such as the search and rescue in dangerous environments \cite{hudson2021heterogeneous}.
The promising solution is 
the so-called feedback motion planning (FMP) strategy that uses feedback (realtime interaction with environments) to endow mobile robots with adaptability towards dynamically changing environments \cite{lavalle2006planning,jaffar2022pip,majumdar2017funnel}.
However, the safety (collision avoidance) problem is often ignored in
current FMP related works \cite{reist2016feedback,yershov2016asymptotically,agha2014firm,lavalle2001algorithms}, especially considering the prior-unknown environment scenario.
Thereby, we propose a safe feedback motion planning (SFMP) strategy to realize safe execution in prior-unknown environments and accomplish given tasks.
We exploit instantaneous local sensory data to stimulate computationally cheap SFMP strategies in prior-unknown environments; Rather than firstly conducting a computationally intensive mapping process and then planning on the constructed map to offer a safe solution.
Besides, the utilized instantaneous local sensory data endows the resulting SFMP strategies with flexibility towards diverse environments.

\subsection{Related Works}
The traditional solutions to the motion planning problem mainly include grid-based \cite{claussmann2019review}, 
sampling-based \cite{lavalle2001rapidly},
and numerical optimization based \cite{tordesillas2021faster} algorithms.
Note that it is hard to offer a complete review due to the page limit.
Thus, the authors pick up the representative works here.
The solutions mentioned above cannot be easily applied to prior-unknown environments given the following two reasons.
Firstly, the effectiveness and performance of the above mentioned methods \cite{claussmann2019review,lavalle2001rapidly,tordesillas2021faster} rely on a pre-built perfect map. This is unavailable for the  (partially) unknown environment scenario.
Secondly, the open-loop motion planning strategies (a function of initial states only) in the works \cite{claussmann2019review,lavalle2001rapidly,tordesillas2021faster}  are not competent to adapt to varying environments, or even small deviations from the expectations in practical applications.
Departing from the mechanism of the above traditional solutions, we utilize realtime interaction with environments to stimulate FMP strategies with resilience towards prior-unknown environments.

To further operate safely in prior-unknown environments, the mobile robot needs to discover and react to potential collisions.
Rather than using the common collision avoidance tools such as artificial potential method \cite{rimon1992exact}, collision cone \cite{sunkara2019collision},  navigation function \cite{tanner2012multiagent}, funnel \cite{jaffar2022pip,majumdar2017funnel}, and reachable set \cite{mitchell2005time}, 
we prefer to use the mechanism of control barrier function (CBF) \cite{ames2016control} to facilitate the SFMP strategy given its simplicity (easier collision check) and rigorousness (theoretical guarantee of safety).
Normally, CBFs are constructed using obstacle information such as location, shape, and number \cite{ames2016control}.
However, complete knowledge of obstacles in an unstructured environment is usually unavailable.
Thus, the online learning of obstacle related CBFs is required if practitioners want to use CBFs to enforce safety in prior-unknown environments.
The neural network parameterized CBFs are learned using a cost function that characterizes essential proprieties of CBFs \cite{zhao2020synthesizing,jin2020neural,jagtap2020control}.
The offline learning of barrier functions using expert demonstrations is adopted in \cite{saveriano2019learning,robey2020learning}.
Besides, the CBF learning is formulated as a classification problem in \cite{srinivasan2020synthesis}, wherein a complete obstacle boundary is identified via the support vector machine method.
The CBF learning problem is solved through a global perspective in the works \cite{zhao2020synthesizing,jin2020neural,jagtap2020control,saveriano2019learning,robey2020learning,srinivasan2020synthesis} mentioned above. 
Alternatively, we attempt to learn CBFs from a local perspective in favour of computation efficiency.
The resulting instantaneous local control barrier functions  (IL-CBFs) are robust to previously-unobserved environments.
Along with the safety problem discussed above, the reaching task to a predetermined goal position can be realized by control Lyapunov function (CLF) based analytical or numerical solutions \cite{freeman1995robust} that are favored with theoretical convergence guarantees to target positions. 
However, either predetermined \cite{freeman1995robust} or learned \cite{abate2020formal} CLF based solutions are inefficient to complete long-horizon tasks in practice.
We solve this problem through a divide-and-conquer approach by discovering subgoals incrementally using sensory data and further constructing associated goal-driven control Lyapunov functions (GD-CLFs) for each subtask.

CBFs are often used with CLFs under a quadratic programming (QP) optimization \cite{capelli2022passivity}.
However, the resulting QP is susceptible to infeasibility; especially considering limited motion commands.
This gap has been marginally considered in existing works.
A promising work \cite{xiao2021high} improves the QP feasibility by using the developed penalty and parameterization methods. 
Besides, control-sharing CBFs \cite{xu2018constrained} and control-sharing CLFs \cite{grammatico2013control} are investigated separately to improve the QP feasibility within consideration of multiple CBF or CLF constraints. 
However, the theoretically promising conclusions in \cite{xu2018constrained, grammatico2013control} no longer hold when CBFs and CLFs are used together.
This work enhances the QP feasibility by enlarging the admissible control spaces (ACSs) of the IL-CBF constraints via our formulated linear programming (LP) optimization,
and introducing a relaxation variable for the GD-CLF constraints similar to \cite{xiao2021safety}.
\subsection{Contributions}
The main contribution of this paper is learning IL-CBFs and GD-CLFs from sensory data to encode safety and task requirements, respectively.
These online learned constraints considered in the QP optimization process allow us to analyze and fulfill requirements of safety and task (convergence to goal positions).
Another contribution is conducting an optimization over the ACSs of IL-CBF constraints to enhance the solution feasibility of the associated QP.
The feasibility of QP under multiple constraints remains an open problem \cite{xiao2021high}. 
Towards this end, we reinvestigate the learned IL-CBF constraints in a motion control space (the axis is the motion command). 
This allows us to design a metric to quantify the volume of the ACS of the IL-CBF and further enlarge its area under an LP optimization process.

The remainder of this paper is organized as follows. Section \ref{sec problem formulation} presents the problem formulation. 
Then, the IL-CBF and GD-CLF learning processes are clarified in Section \ref{sec IL-CBF} and Section \ref{sec GD-CLF}, respectively.
Thereafter, the learned IL-CBFs and GD-CLFs are united through the QP in Section \ref{sec QP problem},
and the strategy to enhance the QP feasibility is shown in Section \ref{sec enlarge ACS}.
The SFMP strategy is numerically validated in Section \ref{sec simu}.
Finally, the conclusion is shown in Section \ref{sec conclusion}.

\emph{Notations:}
Throughout this paper, $\mathbb{R}$, $\mathbb{R}^{+}$, and $\mathbb{R}^{+}_{0}$ denote the set of real, positive, and non-negative real numbers, respectively; 
$\mathbb{N}^{+}$ denotes the set of non-negative integers;
$\mathbb{R}^{n}$ is the Euclidean space of $n$-dimensional real vector; 
$\mathbb{R}^{n \times m}$ is the Euclidean space of $n \times m$ real matrices; 
The $i$-th entry of a vector $x = [x_{1},...,x_{n}]^{\top}\in \mathbb{R}^{n}$ is denoted by $x_i$, and $\left\| x \right\| = \sqrt{\sum_{i=1}^{N}|x|^2}$ is the Euclidean norm of the vector $x$;
The $ij$-th entry of a matrix $D \in \mathbb{R}^{n \times m}$ is denoted by $d_{ij}$, and $\left\|D\right\| = \sqrt{\sum_{i=1}^{n}\sum_{j=1}^{m}|d_{ij}|^2}$ is the Frobenius norm of the matrix $D$. 
For notational brevity, time dependence is suppressed without causing ambiguity.

\section{Problem Formulation} \label{sec problem formulation}
This work investigates the safe operation problem of a mobile robot in previously unforeseen environments. We model the investigated mobile robot as:
\begin{equation} \label{eq control afine sys}
 \begin{bmatrix}
            \dot{p} \\
            \dot{v} \\
        \end{bmatrix}=
         \begin{bmatrix}
            0_{2 \times 2} & I_{2 \times 2} \\
            0_{2 \times 2} & 0_{2 \times 2} \\
        \end{bmatrix}\begin{bmatrix}
            p \\
            v \\
        \end{bmatrix}+
         \begin{bmatrix}
            0_{2 \times 2} \\
            I_{2 \times 2} \\
        \end{bmatrix} u,
\end{equation}
where $p:= [x,y]^{\top}$, $v:=  [v_{x},v_{y}]^{\top}$,  and $u:=  [u_{x},u_{y}]^{\top} \in \mathbb{R}^2$ are the positions, velocities, and motion commands, respectively. 
For simplicity, we assume that the robot localization is perfect, i.e., the accurate vehicle state is available \footnote{
The localization is realizable by the low-cost dead reckoning method. 
Dealing with its cumulative error is a different research direction, which is beyond the scope of this paper.}. 

Assume that there exist multiple prior unknown obstacles $\mathcal{O}_l$ in
an environment $\mathcal{E}$, where $l \in \mathcal{L} := \left\{l|l=1,2,\cdots,L\right\}$ and $L \in \mathbb{N}^+$ is an uncertain value. 
The objective is to design an SFMP strategy $u$ to drive the mobile robot \eqref{eq control afine sys} to operate safely in a prior-unknown environment $\mathcal{E}$ and finally reach the predetermined target position $p_d:= [x_d, y_d]^{\top} \in \mathbb{R}^{2}$. 
We formulate the safe operation problem mentioned above as a constrained optimization problem stated as:
\begin{subequations}\label{eq constraint optimization}
\begin{align}
    \min_{u} J & := \int_{t_0}^{t_f} u^{\top}u \, dt  \label{CO cost}\\  
\st \, \, \, &  \eqref{eq control afine sys} \notag \\  
     & p(t_0)= p_0; \, \, v(t_0)= v_0 \label{initial conditions} \\ 
        & u(t) \in \mathcal{U}, \, \, \forall t \in \left[t_0, t_f \right]\label{CO input constraint} \\
          & p(t) \cap \bigcup_{l=1}^{L} \mathcal{O}_l = \emptyset, \, \, \forall t \in \left[t_0, t_f \right] \label{CO state constriant} \\
     & \left\| p \left(t_f\right)  - p_d \right\| \leq \delta, \label{CO reach task} 
\end{align}
\end{subequations}
where the kinodynamic constraint \eqref{eq control afine sys} and the bounded input space $\mathcal{U} \subseteq \mathbb{R}^2$ in \eqref{CO input constraint} are considered to ensure the resulting SFMP strategy obeying the physical feasibility.
A prior set threshold $\delta \in \mathbb{R}^+$ in \eqref{CO reach task} is used to check whether the reach task is completed.
A quadratic energy function is adopted in \eqref{CO cost} to reflect designers' preference for the energy minimization.

\begin{figure}[!t]
    \centering
    \includegraphics[width=3.4in]{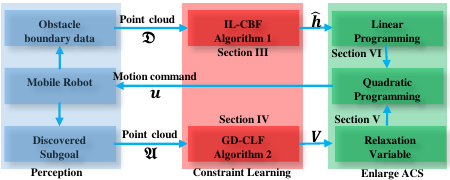}
    \caption{Schematic of the SFMP strategy that maps raw sensory data to motion commands.
    The IL-CBFs learned from sensory data in Section \ref{sec IL-CBF} characterize the obstacle boundaries;
    The decomposed short-horizon subtasks are encoded by GD-CLFs clarified in Section \ref{sec GD-CLF};
    The LP optimization is conducted to enlarge the ACSs in Section \ref{sec enlarge ACS} to improve the feasibility of the QP formulated in Section \ref{sec QP problem}.
    }
    \label{fig schematic of structure}
\end{figure}

The aforementioned safe operation problem \eqref{eq constraint optimization} is nontrivial given 
the constraints indicating different (might conflicting) objectives of safety and performance maximization;
and the requirement of constraint satisfaction under uncertainty (limited knowledge of the environment $\mathcal{E}$).
This work develops an SFMP strategy to solve \eqref{eq constraint optimization}, whose mechanism is illustrated in Fig. \ref{fig schematic of structure}.
In particular, we use perception inputs to learn IL-CBFs and GD-CLFs that are utilized to achieve collision avoidance and accomplish given tasks.

\section{IL-CBF Online Learning} \label{sec IL-CBF}
This section elucidates the mechanism of learning IL-CBFs from sensory data. 
In particular, the detected local obstacle information is utilized to learn local barrier functions to describe the partial obstacle boundaries; and the learned local barrier functions update along with continuously coming data to tackle the prior-unknown environment.
Our developed IL-CBFs are employed to formulate the QP problem in Section \ref{sec QP problem} to conduct collision avoidance with prior-unforeseen obstacles.

\begin{figure}[!t]
    \centering
    \includegraphics[width=3.4in]{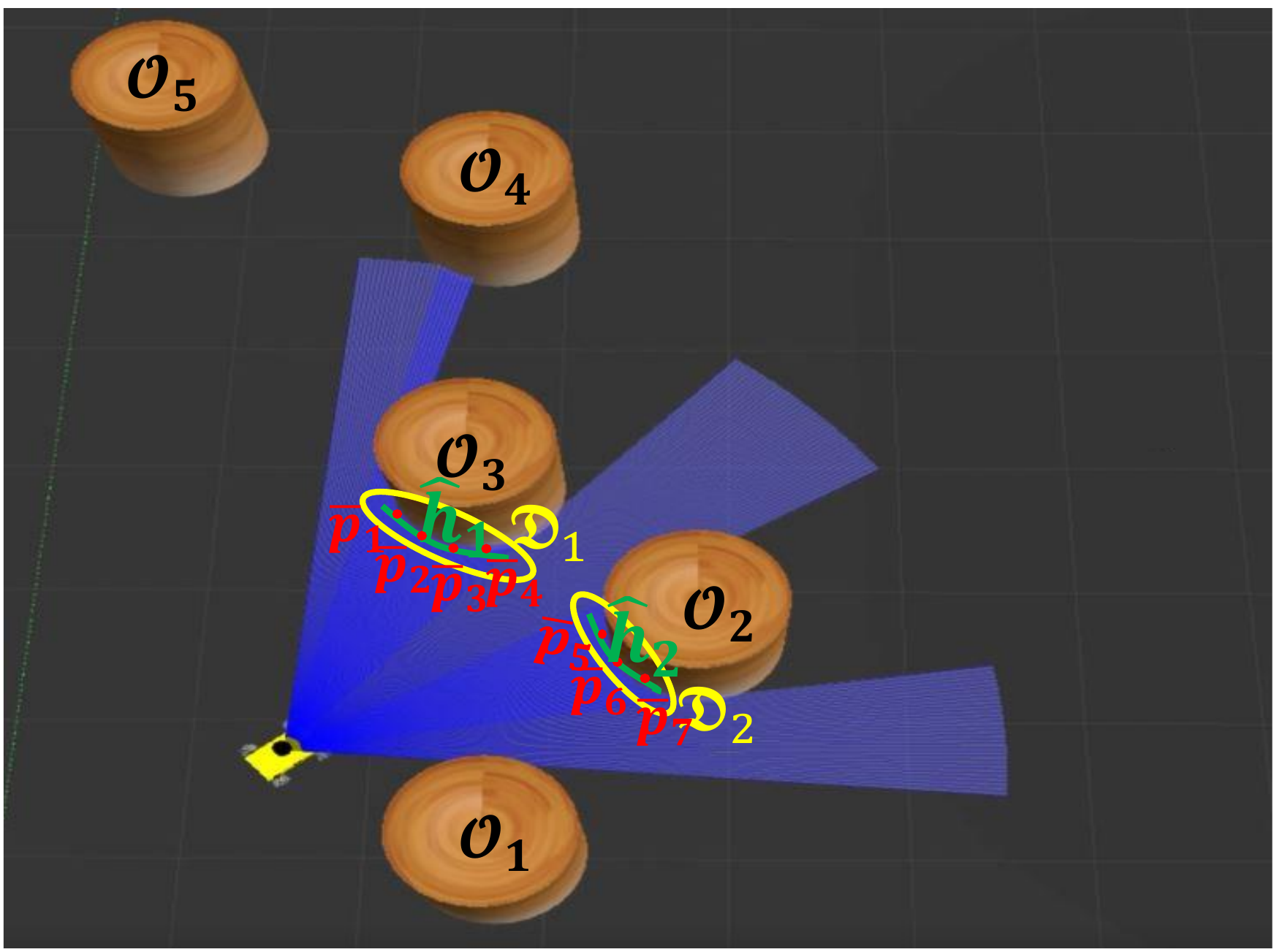}
    \caption{Graphical illustration of IL-CBFs and obstacles.
    The whole boundaries of obstacles $\mathcal{O}_l$ are  described by explicit CBFs $h_l = (x - x_{o_{l}})^2 + (y - y_{o_{l}})^2 - r^2_l$, $c_l = \left(x_{o_{l}}, y_{o_{l}}\right) $, $l = 1, 2,3,4,5$.
    The mobile robot observes $\mathfrak{D} = \left\{\bar{p}_1, \bar{p}_2, \cdots, \bar{p}_{7} \right\}$ and classifies $\mathfrak{D}$ into subgroups $\mathfrak{D}_k$, $k = 1,2$. Thus, $K = 2$, and $I_1 = 4$, $I_2 = 3$ here.
    The mobile robot learns $\hat{h}_k$ based on the $k$-th data subgroup $\mathfrak{D}_k$.
   }
    \label{fig mechanism of ILCBF}
\end{figure}

As illustrated in Fig. \ref{fig mechanism of ILCBF}, the whole boundaries of the obstacles $\mathcal{O}_l$ in $\mathcal{E}$ could be described by the barrier functions $h_l(p) \in \mathbb{R}$ using the complete knowledge of obstacles \cite{ames2016control}, which is however unavailable in our investigated problem \eqref{eq constraint optimization}. 
Thus, the explicit forms of $h_{l}(p)$ that characterize the dangerous regions $\mathcal{O}_l$ are unavailable. 
We observe in Fig. \ref{fig mechanism of ILCBF} that only partial obstacle boundaries of $\mathcal{O}_l$ pose threats to the mobile robot safety at a certain period. 
This motivates us to utilize the local sensory data to learn the local barrier functions, corresponding to the partial obstacle boundary within the mobile robot's sensor horizon, to address the collision avoidance problem. 

Assume that the mobile robot is embedded with a sensor with a restricted angle $S_{\theta}$ and a limited horizon $S_r$. 
The value of $S_{\theta}$ is given, and the value of $S_r$ satisfies
\begin{equation}\label{eq sr}
    S_r \geq  D_{\text{brake}} := \left\|v_{\max}\right\|^2/ \left\|a_{\max}\right\|,
\end{equation}
where $v_{\max}$, $a_{\max} \in \mathbb{R}^2$ are the maximum velocity and breaking acceleration of the mobile robot \eqref{eq control afine sys}. 
Here $D_{\text{brake}}$ denotes the travelled distance when the mobile robot in the maximum velocity brakes using the maximum breaking acceleration. 
\begin{remark}
The setting of the sensor horizon $S_r$ in \eqref{eq sr} is beneficial to the emergence case where our developed SFMP strategy fails to guarantee safety. In this scenario, the mobile robot brakes to avoid collisions.
\end{remark}
The sensor provides a point cloud $\mathfrak{L}$. 
We term 
$\mathfrak{D} := \left\{\bar{p}_1, \bar{p}_2, \cdots\right\} \subset \mathfrak{L}$ 
as the data group of the sensed obstacle boundaries, wherein $ \bar{p}_i:=  [\bar{x}_{{{i}}},\bar{y}_{{{i}}}]^{\top} \in \mathbb{R}^2$ is the position of the $i$-th detected obstacle boundary point.
In an environment $\mathcal{E}$ with densely populated obstacles, data points in $\mathfrak{D}$ might concern multiple isolated obstacles, as displayed in Fig. \ref{fig mechanism of ILCBF}. Therefore, we adopt the robust clustering algorithm--\emph{density-based spatial clustering of applications with noise (DBSCAN)} \cite{ester1996density}--to cluster $\mathfrak{D}$ into multiple subgroups $\mathfrak{D}_{k} :=  \left\{\bar{p}_{k_{1}}, \bar{p}_{k_{2}}, \cdots \right\}$, wherein
$\bar{p}_{k_{i}}:=  [\bar{x}_{k_{i}},\bar{y}_{k_{i}}]^{\top} \in \mathbb{R}^2$ denotes the $i$-th data point of the $k$-th data subgroup $\mathfrak{D}_{k}$, $i \in \mathcal{I}:=  \left\{i|i = 1, \cdots, I_k\right\}$ with  $I_k \in \mathbb{N}^+$ being the total number of data points in the  $\mathfrak{D}_{k}$,
and $k \in \mathcal{K}:=  \left\{k|k = 1, \cdots, K\right\}$ with $K \in \mathbb{N}^+$ being the sum of the local obstacle boundary considered in the current period.

\begin{remark}
The DBSCAN algorithm is compatible with our IL-CBF learning process given that it could determine the number of to be learned IL-CBFs (i.e., the values of $K$)  automatically without using prior knowledge of environments.
\end{remark}

In the following, we clarify the mechanism of the IL-CBF learning focusing on the $k$-th data subgroup $\mathfrak{D}_{k}$.
Assume that $i$-th data pair $\bar{p}_{k_{i}}$ satisfies
\begin{equation}\label{LR part}
    \bar{y}_{{k}_{i}}=  \mathcal{F}(\bar{x}_{{k}_{{i}}},\zeta_{k}) + \varepsilon_{k}, \, \, k \in \mathcal{K},
\end{equation}
where $\mathcal{F}(\bar{x}_{{k}_{{i}}},\zeta_k) \in \mathbb{R}$ is one $n$-th degree polynomial function with a parameter $\zeta_k \in \mathbb{R}^{n+1}$ to be learned;
and $\varepsilon_{k} \sim  N(0,\sigma^2)$ denotes an assumed  Gaussian sensor noise with a zero mean and a constant variance $\sigma \in \mathbb{R}$.

\begin{remark}\label{choice of approximation scheme}
There exist multiple choices for $\mathcal{F}$, such as Gaussian models, linear fitting, and rational polynomials \cite{rawlings2001applied}. Considering the generality and simplicity issues, a polynomial model is chosen here.
\end{remark}
Based on \eqref{LR part} and the point cloud $\mathfrak{D}_{{k}}$ from the sensor, $\zeta_{k}$ is learned to minimize the approximation error
\begin{equation}\label{LR}
    \hat{\zeta}_{k} = \arg \min_{\zeta_{k}} \sum_{i=1}^{I_k} \left(\bar{y}_{{k}_{{i}}}-\mathcal{F}(\bar{x}_{{k}_{{i}}},\zeta_{k})\right)^2, \, \, k \in \mathcal{K}. 
\end{equation}  
To address potential noises and outliers that exist in the measurement data, the robust regression technique--\emph{M-estimate} \cite{holland1977robust}--is adopted here.
By using the M-estimate, the learning of $\zeta_{k}$ in \eqref{LR} is rewritten as 
\begin{equation}\label{IRLS}
    \hat{\zeta}_{k} = \arg \min_{\zeta_{k}} \sum_{i=1}^{I_{k}} \rho \left(\frac{\bar{y}_{{k}_{{i}}}-\mathcal{F}(\bar{x}_{{k}_{{i}}},\zeta_{k})}{\gamma}\right),  \, \, k \in \mathcal{K},
\end{equation}
where $\rho(r) = c^2/(1-(1-(r/c)^2)^3)$ is a robust loss function with $c = 1.345$;
$\gamma$ is a scale parameter estimated as
$\gamma = 1.48 \left[ \text{med}_{i} \left| \left(\bar{y}_{k_{i}}-\mathcal{F}(\bar{x}_{k_{i}},\zeta_{k_{0}}) \right) - \text{med}_{i} \left(\bar{y}_{k_{i}}-\mathcal{F}(\bar{x}_{k_{i}},\zeta_{k_{0}}) \right) \right| \right]$ with $\zeta_{k_{0}}$ being the initial value of $\zeta_{k}$.
More details about the \emph{M-estimate} approach are referred to \cite{holland1977robust}.

Using the learned $\hat{\zeta}_{k}$ \eqref{IRLS}, we construct the IL-CBF $\hat{h}_k$ as
\begin{equation}\label{eq IL-CBF}
    \hat{h}_k = y-\mathcal{F}(x,\hat{\zeta}_{k}), \, \, k \in \mathcal{K}.
\end{equation}

The IL-CBF learning process mentioned above is summarized in Algorithm \ref{LCBF learn algorithm}. 
The mobile robot uses Algorithm \ref{LCBF learn algorithm} to update the learned IL-CBFs continuously based on the newly observed sensory data during the operation process. 
The IL-CBF learning is favored with computation simplicity. 
Thus, it is practical to update the learned IL-CBFs at each step, which is favourable for the mobile robot to observe the environmental changes in time and make corresponding reactions.
\begin{algorithm}[!t]
        \caption{IL-CBF Online Learning Algorithm}
        \label{LCBF learn algorithm}
        \begin{algorithmic}[1] 
            \Require Point cloud $\mathfrak{D}$;
            \Ensure $\hat{h}{_k}, k = 1, \cdots, K $; 
            \State $K =$ DBSCAN $ (\mathfrak{D})$ \Comment{Robust clustering}
            \For {$k = 1:K$}
                \State $\hat{\zeta}_{k} =$ M-estimate $(\mathfrak{D}_{k})$ \eqref{IRLS} \Comment{Robust regression}
                \State $\hat{h}_k = y-\mathcal{F}(x,\hat{\zeta}_{k})$ \eqref{eq IL-CBF} 
            \EndFor
            \end{algorithmic}
\end{algorithm}
\begin{remark}
Alternatively, we are able to realize the CBF learning in an incremental way along with a steady stream of data, i.e., attempting to gradually learn one global barrier function that describes the whole obstacle boundary. 
However, we found in practice that this increment learning approach shows no obvious advantage in terms of collision avoidance but introduces additional computation loads. Thus, we forgo using all detected data to gradually build a perfect map, rather than only using instantaneous local sensory information.
\end{remark}

\begin{remark}
The clarified IL-CBF learning in this section is especially compatible with low-end sensors that only provide low-dimensional data.
The limited data, however, is not enough to build a global map or describe the whole obstacle boundary.
\end{remark}

\section{GD-CLF Automatic Construction} \label{sec GD-CLF}
The data group $\mathfrak{D}$ concerning the detected obstacle boundaries is utilized in Section \ref{sec IL-CBF} to facilitate collision avoidance in prior-unknown environments. 
This section exploits the remaining local collision-free sensory data group $\mathfrak{A}:= \mathfrak{L} \ominus \mathfrak{D}$ to complete the long-horizon task. 
Specifically, we first utilize the data group $\mathfrak{A}$ to discover subgoals using a Euclidean distance metric.
Then, we construct the associated GD-CLF for each subtask (subgoal).
The automatically constructed GD-CLFs serve as constraints of the QP optimization in Section \ref{sec QP problem}, whose solution ensures that the mobile robot travels toward the discovered subgoals incrementally and finally reaches the destination.

\begin{figure}[!t]
    \centering
    \includegraphics[width=3.4in]{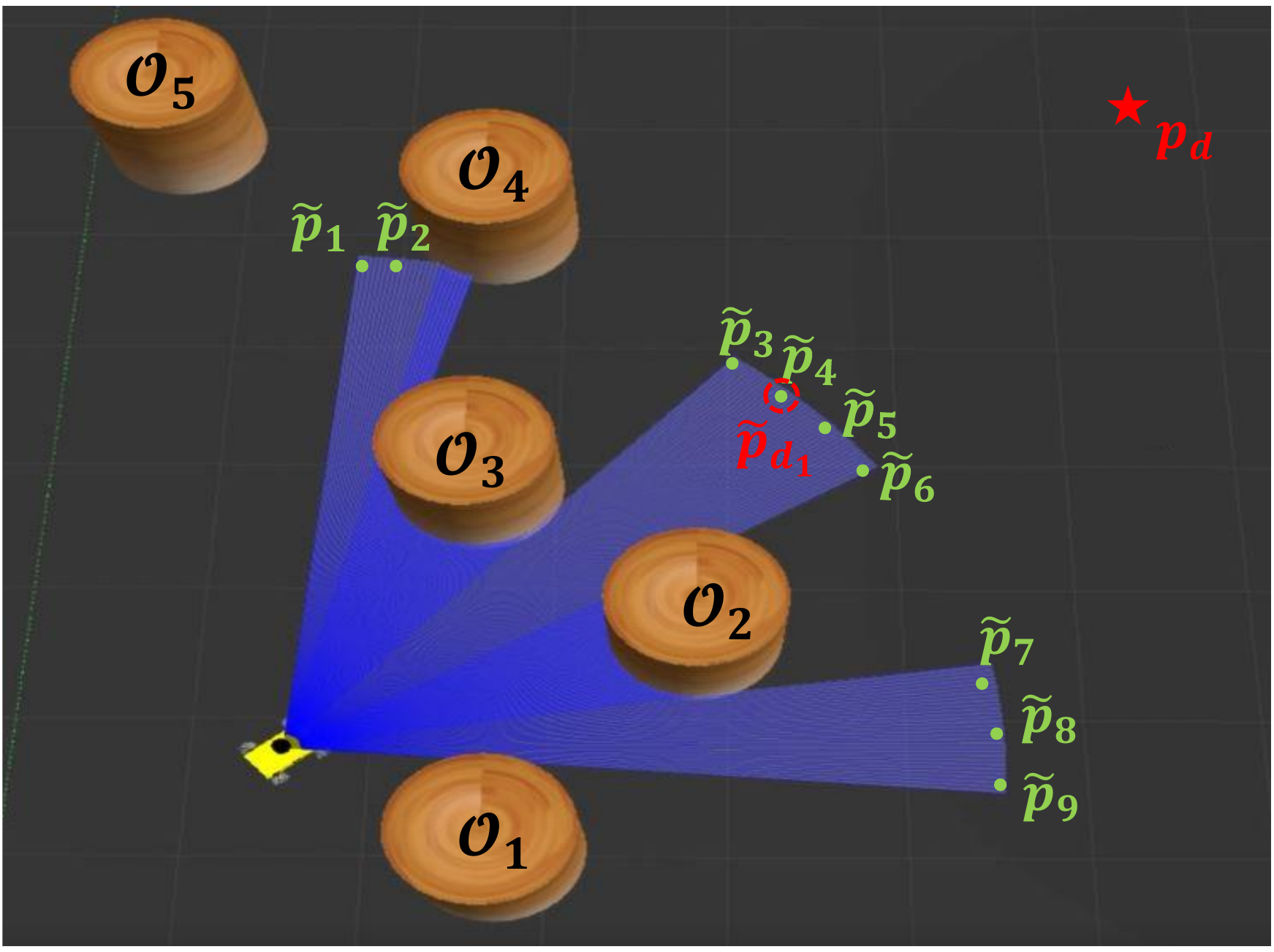}
    \caption{Graphical illustration of GD-CLFs and subgoals.
    The mobile robot uses the collision-free data group $\mathfrak{A} = \left\{\tilde{p}_1, \tilde{p}_2, \cdots, \tilde{p}_{9} \right\}$ and Algorithm \ref{GLCLF learn algorithm} to determine the position $\tilde{p}_4 \in \mathfrak{A}$ as its first subgoal $\tilde{p}_{d_{1}}$. 
    Then, the constructed GD-CLF $V_1$ guides the robot toward $\tilde{p}_{d_{1}}$.
    The robot would determine its $j+1$-the subgoal when it arrives at $\delta$-neighboured around the $j$-th subgoal.
    Following the above-mentioned process, the mobile robot would travel along the successively discovered subgoals $\tilde{p}_{d_{2}}$, $\tilde{p}_{d_{3}}$, $\cdots$, and finally reaches the goal $p_d$.  
    }
    \label{fig method GD-CLF}
\end{figure}

Normally, the common CLF used in \cite{freeman1995robust,abate2020formal} is inefficient to account for a long-horizon goal.
Thus, through a divide-and-conquer perspective, we use sensory data $\mathfrak{A}$ to discover the subgoals $\tilde{p}_{d_{j}}:=  [x_{d_{j}},y_{d_{j}}]^{\top} \in \mathbb{R}^2$,  $j \in \mathcal{J}:=  \left\{j|j = 1, \cdots, J \right\}$ with $J \in \mathbb{N}^+$, based on a Euclidean distance metric (line 3 of Algorithm \ref{GLCLF learn algorithm}). 
In particular, we choose the nearest collision-free waypoint ( $\tilde{p}_{4}$  in Fig. \ref{fig method GD-CLF} for example) toward the goal position $p_d$ as the next subgoal ($\tilde{p}_{d_{1}}$ in Fig. \ref{fig method GD-CLF} for example).
The automatically determined intermediate waypoints $\tilde{p}_{d_{j}}$ would forwardly progress toward the final desired position $p_d$.

\begin{algorithm}[!t]
        \caption{GD-CLF Online Learning Algorithm} 
        \label{GLCLF learn algorithm}
        \begin{algorithmic}[1] 
            \Require Point cloud $\mathfrak{A}:= \left\{\tilde{p}_1, \tilde{p}_2, \cdots\right\}$; Robot position $p$.
            \Ensure $\tilde{p}_{d_{j}}$, and $V_{j}$, $j = 1, \cdots, J $;
            \State $\tilde{p}_{d_{1}} = \arg \min_{\tilde{p}_{i} \in \mathfrak{A}} \left\| \tilde{p}_{i} - p_d \right\|$ and get $V_{1}$ \eqref{eq GLCLF}
            \If {$\left\| p - \tilde{p}_{d_{j}} \right\| \leq \delta$} 
                \State $\tilde{p}_{d_{j}} = \arg \min_{\tilde{p}_{i} \in \mathfrak{A}} \left\| \tilde{p}_{i} - p_d \right\|$
                \State $j = j+1$ and update $V_{j}$ \eqref{eq GLCLF}
            \EndIf
            \end{algorithmic}
\end{algorithm}

The automatically determined subgoals $\tilde{p}_{d_{j}}$ from Algorithm \ref{GLCLF learn algorithm} divide the long-horizon task into $J$ short-horizon subtasks. For each subtask, we construct the GD-CLF :
\begin{equation} \label{eq GLCLF}
    V_j = (p-\tilde{p}_{d_{j}})^{\top} P (p-\tilde{p}_{d_{j}}) + (v-v_{dj})^{\top} Q (v-v_{dj}), j \in \mathcal{J}
\end{equation}
where $P$, $Q \in \mathbb{R}^{2 \times 2}$ are predetermined positive definite matrices; and $v_{d_{j}} \in \mathbb{R}^2$ could be a zero or a prior-given constant velocity vector.
The constructed GD-CLF $V_j$ \eqref{eq GLCLF} updates as the subgoal $\tilde{p}_{d_{j}}$ refreshes using Algorithm \ref{GLCLF learn algorithm}.
\begin{remark}
Note that we construct IL-CBFs \eqref{eq IL-CBF} in Section \ref{sec IL-CBF} and GD-CLFs \eqref{eq GLCLF} in Section \ref{sec GD-CLF} assuming that $\mathcal{U} = \mathbb{R}^2$ for convenience, i.e., the influence of input saturation is ignored temporally.
This problem is later tackled in Section \ref{sec enlarge ACS} by explicitly analysing the potential conflicts between IL-CBF, GD-CLF, and input constraints.
\end{remark}
 \section{Safe Feedback Motion Planning Strategy} \label{sec QP problem}
This section incorporates the learned IL-CBFs \eqref{eq IL-CBF} and the constructed GD-CLFs \eqref{eq GLCLF} in a QP optimization to generate the SFMP strategy that drives the mobile robot to safely reach the target position incrementally.

By dividing the period $[t_0, t_f]$ into multiple intervals $[t_0+mT,t_0+(m+1)T]$ \cite{xiao2021high}, where $m \in \mathbb{N}^{+}$, and $T \in \mathbb{R}^+$ is the sampling time, 
we reformulate the original safe operation problem  \eqref{eq constraint optimization} into a sequence of QPs at each interval:
\begin{subequations}\label{eq QP}
\begin{align}
    \min_{u,\nu}  &  \, \, u(t)^{\top}u(t) + \bar{c}_1 \nu^2(t) \label{QP cost}\\
\st  
    & \, \, \eqref{eq control afine sys}, \, \, \eqref{initial conditions}, \, \, \eqref{CO input constraint} \notag \\
     & \, \, \Ddot{\hat{h}}_k + \alpha_{k_{1}} \dot{\hat{h}}_k + \alpha_{k_{2}} \hat{h}_k \geq 0, \label{QP CBF}\\
     &  \, \, \dot{V}_j + \bar{c}_2 V_j \leq \nu,\label{QP reach task} 
\end{align}
\end{subequations}
where $\nu(t) \in \mathbb{R}$ is a relaxation variable to relax the  GD-CLF constraint to improve the QP feasibility \cite{xiao2021safety};
$\alpha_{k_{1}}$, $\alpha_{k_{2}}$, $\bar{c}_1$, $\bar{c}_2 \in \mathbb{R}$ are parameters to be determined.
The reformulated QP problem \eqref{eq QP} unifies the safety requirements \eqref{CO input constraint}, \eqref{QP CBF}, the task requirements \eqref{QP reach task}, and the optimization over energy \eqref{QP cost} to generate a multi-objective SFMP strategy that drives the mobile robot to progressively reach subgoals while avoiding obstacles.  
Note that our developed SFMP strategy from \eqref{eq QP} only requires the information of the mobile robot position $p$ and the target position $p_d$ to solve the safe operation problem \eqref{eq constraint optimization} in prior-unknown environments.
\begin{remark}
Although the proposed SFMP strategy \eqref{eq QP} is restricted to the mobile robots following a second-integrator-type kinematics \eqref{eq control afine sys},  it could be easily extended to the mobile robots following an unicycle-type kinematics according to the method proposed in \cite{xiao2022control}.
Besides, the specific cost function \eqref{QP cost} in a quadratic input form is utilized for efficient computation concerns. 
This is especially worthwhile for low-cost platforms with limited computational resources.
The online learned IL-CBFs \eqref{QP CBF} and the automatically constructed GD-CLFs \eqref{QP reach task} are not restricted to specific sensors.
The required data could be provided by different sensors such as LiDAR or cameras.
\end{remark}

\section{Optimized Admissible Control Space} \label{sec enlarge ACS}
The potential conflicts between constraints \eqref{CO input constraint}, \eqref{QP CBF}, and \eqref{QP reach task} might result in the infeasibility problem of the QP \eqref{eq QP} formulated in Section \ref{sec QP problem}.
This section formulates an optimization over the ACS of the IL-CBF associated constraint \eqref{QP CBF} to enhance the solution feasibility of QP. 

Denoting the ACSs for constraints \eqref{QP CBF} and \eqref{QP reach task} as $\mathcal{A}_1  :=  \left\{ u \in \mathbb{R}^{2} | \Ddot{\hat{h}}_k + \alpha_{k_{1}} \dot{\hat{h}}_k + \alpha_{k_{2}} \hat{h}_k \geq 0, k \in \mathcal{K} \right\}$, and $\mathcal{A}_2  :=  \left\{ u \in \mathbb{R}^{2} | \dot{V}_j + c_2 V_j \leq \nu \right\}$, respectively.
Thereby, the shared control space concerning constraints  \eqref{CO input constraint}, \eqref{QP CBF}, and \eqref{QP reach task} is termed as $\mathcal{S} = \mathcal{A}_1 \cap \mathcal{A}_2 \cap \mathcal{U}$. 
It is desirable that $\mathcal{S} \ne \emptyset$ always holds, i.e., the feasibility of the QP problem is always guaranteed. 
This is a nontrivial problem; especially multiple constraints are considered.
Improving the possibility of satisfying $\mathcal{S} \ne \emptyset$ is equivalent to enlarging the volume of $\mathcal{S}$.
Given that the relationship between sets $\mathcal{A}_1$ and $\mathcal{A}_2$ is hard to be described and the volume of $\mathcal{U}$ is predetermined, we could transform the enlargement of the volume of $\mathcal{S}$ into the enlargement of the volumes of ACSs $\mathcal{A}_1$ and $\mathcal{A}_2$ independently.
A relaxation variable $\nu$ has been used in \eqref{QP reach task} to enlarge the volume of $\mathcal{A}_2$. 
In the following, we attempt to enlarge the volume of the ACS $\mathcal{A}_1$ to improve the feasibility of the QP problem \eqref{eq QP}.
In particular, we firstly seek for a criterion for the volume of the ACS $\mathcal{A}_1$ in Section \ref{sec criterion of ACS} by 
investigating the relationship between sets $\mathcal{A}_1$ and $\mathcal{U}$.
Then, an LP optimization problem is formulated in Section \ref{sec Optimized ACS} to optimize the above volume criterion to 
enlarge the volume of the ACS $\mathcal{A}_1$.

\subsection{Criterion of ACS} \label{sec criterion of ACS}
The enlargement of the ACS $\mathcal{A}_1$ is equivalent to enlarge each IL-CBF $\hat{h}_k$ associated ACS that is denoted as $\mathcal{A}_{1_{k}} := \left\{ u \in \mathbb{R}^{2} | \Ddot{\hat{h}}_k + \alpha_{k_{1}} \dot{\hat{h}}_k + \alpha_{k_{2}} \hat{h}_k \geq 0 \right\}$, $k \in \mathcal{K}$. 
The explicit form of the learned $k$-th IL-CBF follows $\hat{h}_k = y- \hat{\zeta}_k^{\top}\Phi$, where $\Phi := \left[1, x, x^2, \cdots, x^n \right]$.
We substitute the explicit $\hat{h}_k$ into \eqref{QP CBF} and rewrite the inequality as 
\begin{equation} \label{eq uxuy}
    A u_x + u_y + a^{\top}_k \Psi > 0,
\end{equation}
where $A := \hat{\zeta}_k^{\top} \pdv{\Phi}{x} \in \mathbb{R}$, $\alpha_k = [\alpha_{k_{1}}, \alpha_{k_{2}}]^{\top} \in \mathbb{R}^2$, $\Psi := \left[ \hat{\zeta}_k^{\top} \pdv{\Phi}{x} v_x - v_y, \hat{\zeta}_k^{\top} \pdv[2]{\Phi}{x} v^{2}_x + \hat{\zeta}_k^{\top} \Psi - y \right]^{\top} \in \mathbb{R}^2$.

\begin{figure}[!t]
  \centering
    \subfloat[$A > 0$ case.]{\includegraphics[width=1.8in]{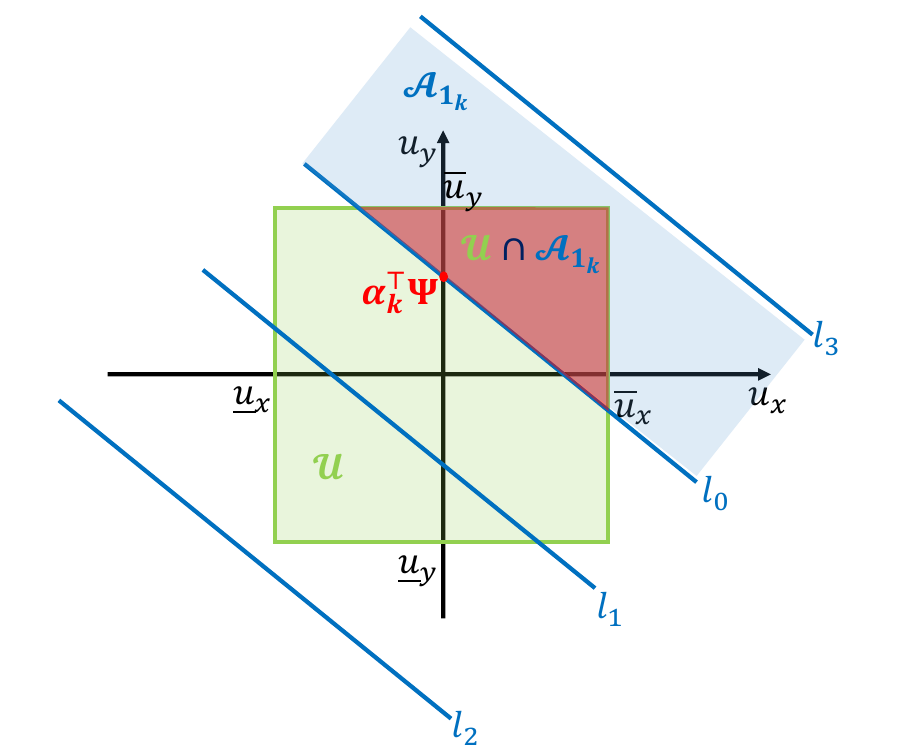}%
    \label{case 1 set}}
    \subfloat[$A < 0$ case]{\includegraphics[width=1.8in]{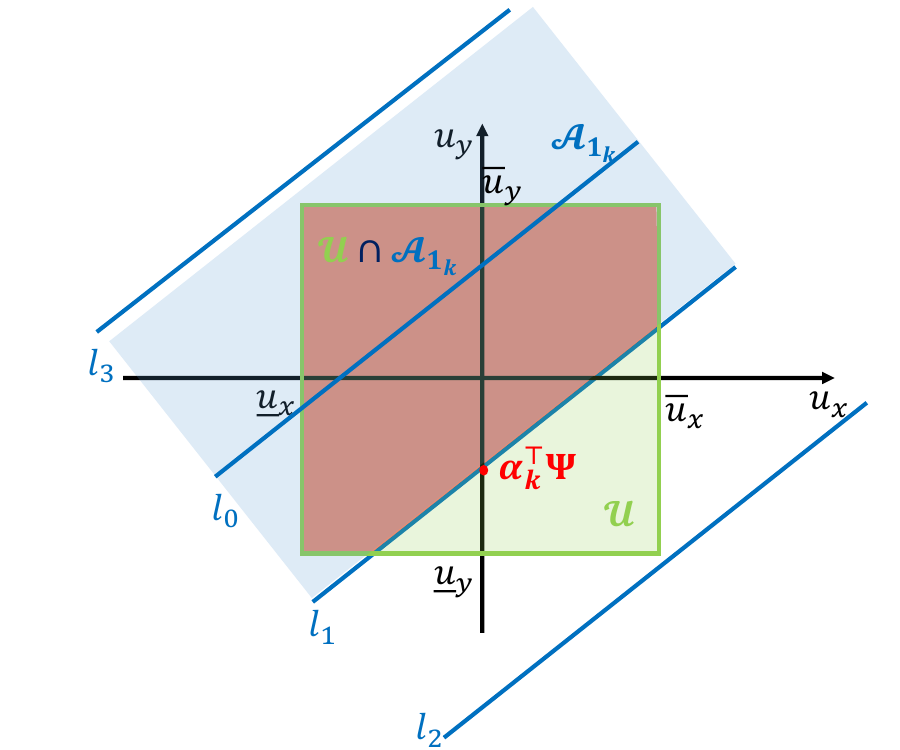}%
    \label{case 2 set}}
    \caption{The geometric interpretation of the sets $\mathcal{A}_{1_{k}}$ (the blue shaded area) and $\mathcal{U}$ (the green shaded area)}. Here $l_k = A u_x + u_y + a^{\top}_k \Psi = 0$. The comparison of the volume of $\mathcal{A}_{1_{k}}$ follows $\mathcal{A}^{l_2}_{1_{k}} > \mathcal{A}^{l_1}_{1_{k}} > \mathcal{A}^{l_0}_{1_{k}} > \mathcal{A}^{l_3}_{1_{k}}$ for both two cases. For the $l_3$ case, $\mathcal{A}_{1_{k}} \cap \mathcal{U} = \emptyset$, i.e., there is no feasible motion command to ensure safety.
    \label{fig ACS and motion command}
\end{figure}

Based on the reformulated \eqref{eq uxuy}, the geometric interpretations of the ACS $\mathcal{A}_{1_{k}}$ as well as the limited motion command set $\mathcal{U}$ are depicted in Fig. \ref{fig ACS and motion command}. 
We found that a smaller value of $a^{\top}_k \Psi$ implies a larger area of the ACS $\mathcal{A}_{1_{k}}$. 
Thus, it is reasonable to choose the value of $a^{\top}_k \Psi$ as a metric to quantify the volume of the ACS $\mathcal{A}_{1_{k}}$, which is optimized in the subsequent subsection. 
\subsection{Optimization of ACS} \label{sec Optimized ACS}
This subsection clarifies the optimization over the metric $a^{\top}_k \Psi$, which is formulated as a LP optimization problem
\begin{subequations}\label{eq optimized ACS}
\begin{align}
    \min_{\alpha_k}  & \, \, \, \, \, \, \alpha^{\top}_k \Psi \label{LP cost}\\
\st  \, \, & 0 < \alpha_{k_{1}}, \alpha_{k_{2}} < \overline{\alpha}_k  \label{LP constraint 1} \\
     &  a^2_{k_{1}} - 4\alpha_{k_{2}} \geq 0 \label{LP constraint 2}
\end{align}
\end{subequations}
where $\overline{\alpha}_k \in \mathbb{R}^+$ is the predetermined bound for the optimization variable. The formulated LP \eqref{eq optimized ACS} is solved by the off-the-self \emph{fmincon} solver.
The core idea of the above LP is to select suitable values of $\alpha_{k_{1}}$ and $\alpha_{k_{2}}$ to minimize $\alpha^{\top}_k \Psi$ while respecting constraints \eqref{LP constraint 1} and \eqref{LP constraint 2}. 
A decreased $\alpha^{\top}_k \Psi$ leads to a enlarged  $\mathcal{A}_{1_{k}}$. Thereby, the QP feasibility is improved.

\begin{remark}
The constraints \eqref{LP constraint 1} and \eqref{LP constraint 2} are the simplification of the following three constraints: (1) $a^2_{k_{1}} - 4\alpha_{k_{2}} \geq 0$; (2) $\frac{-\alpha_{k_{1}}+\sqrt{a^2_{k_{1}} - 4\alpha_{k_{2}}}}{2} < 0$;
(3) $\frac{-\alpha_{k_{1}}-\sqrt{a^2_{k_{1}} - 4\alpha_{k_{2}}}}{2} < 0$.
These three constraints ensure that the roots of \eqref{QP CBF}'s related polynomials 
$\mathcal{P}(\lambda) =  \lambda^2 + \alpha_{k_{1}} \lambda + \alpha_{k_{2}}$ are all negative. 
These constraints ensure that the optimized parameter $\alpha^{*}_k$ leads to valid HO-CBFs.
More details about HO-CBFs are referred to \cite{xu2018constrained}.
\end{remark}

\section{Numerical Simulation} \label{sec simu}
This section conducts numerical simulations to validate the efficiency of our proposed SFMP strategy \eqref{eq QP}. 
In particular, Section \ref{sec sim enlarge ACS} focuses on a reach-avoid benchmark problem to validate the effectiveness of the LP optimization \eqref{eq optimized ACS}. The resulting enlarged ACS leads to better performance.
Then, we validate the efficiency of the SFMP strategy under two representative environments: an obstacle-filled outdoor scenario in Section \ref{sec sim outdoor env}, and a maze indoor scenario in Section \ref{sec sim indoor env}. 
The mobile robot safely operates in the unforeseen outdoor or the maze indoor environment and completes the given long-horizon reach task using the SFMP strategy, generated by solving the QP \eqref{eq QP} within consideration of our developed IL-CBF  \eqref{eq IL-CBF} and GD-CLF \eqref{eq GLCLF}. The QP feasibility is preserved during the operation process via the LP optimization \eqref{eq optimized ACS}.
Furthermore, the effectiveness of our proposed SFMP strategy is validated based on the constructed high-fidelity simulator in Section \ref{sec ros simu},
wherein the repeated validations in one large environment and the comparison with one baseline are considered.

\subsection{Validation of Optimized ACS}\label{sec sim enlarge ACS}
\begin{figure}[h]
 \centering
    \subfloat[The comparison regarding $p(t)$.]{\includegraphics[width=1.8in]{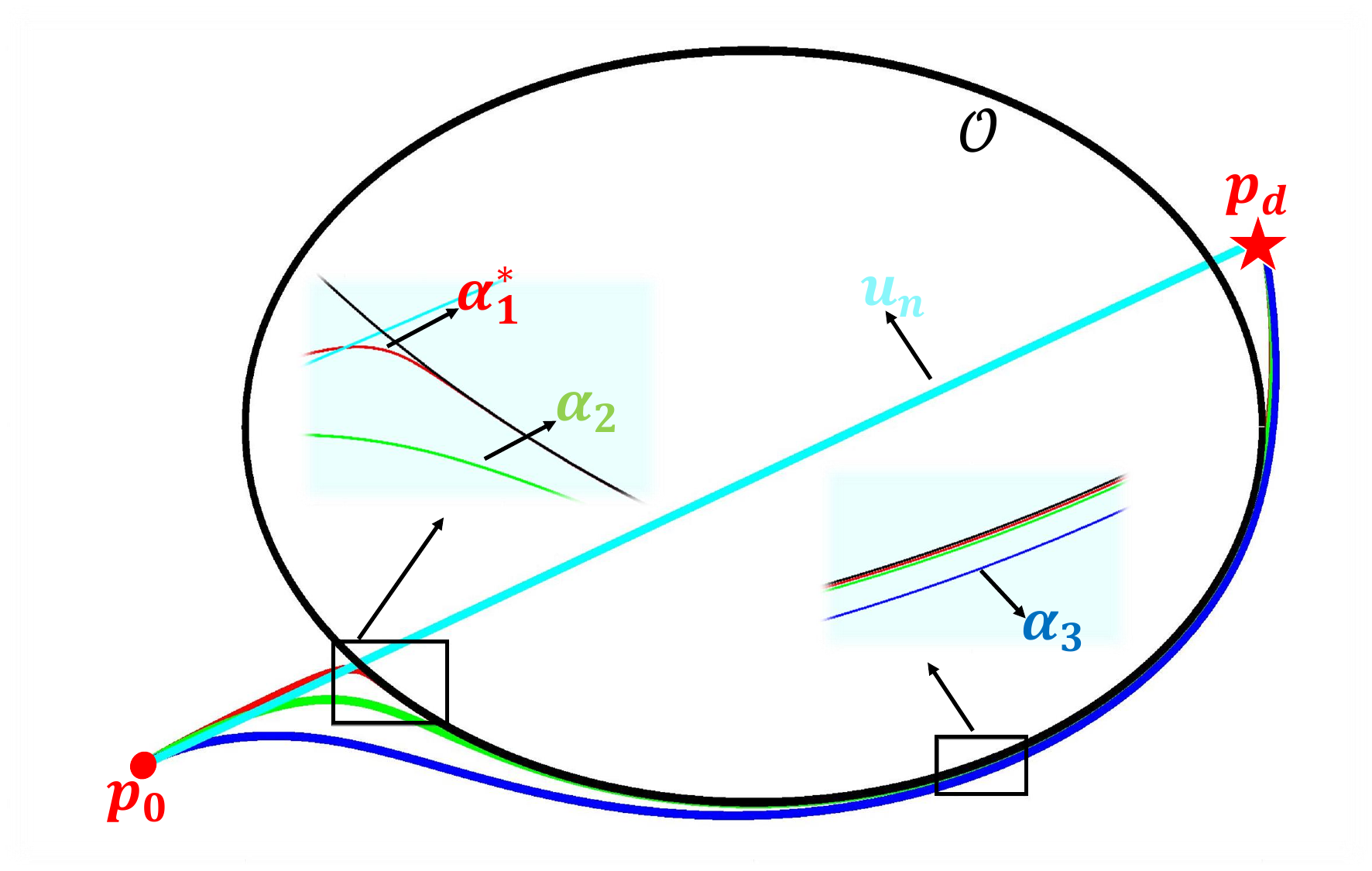}%
    \label{sim pd traj}}
    \subfloat[The comparison regarding ACS.]{\includegraphics[width=1.8in]{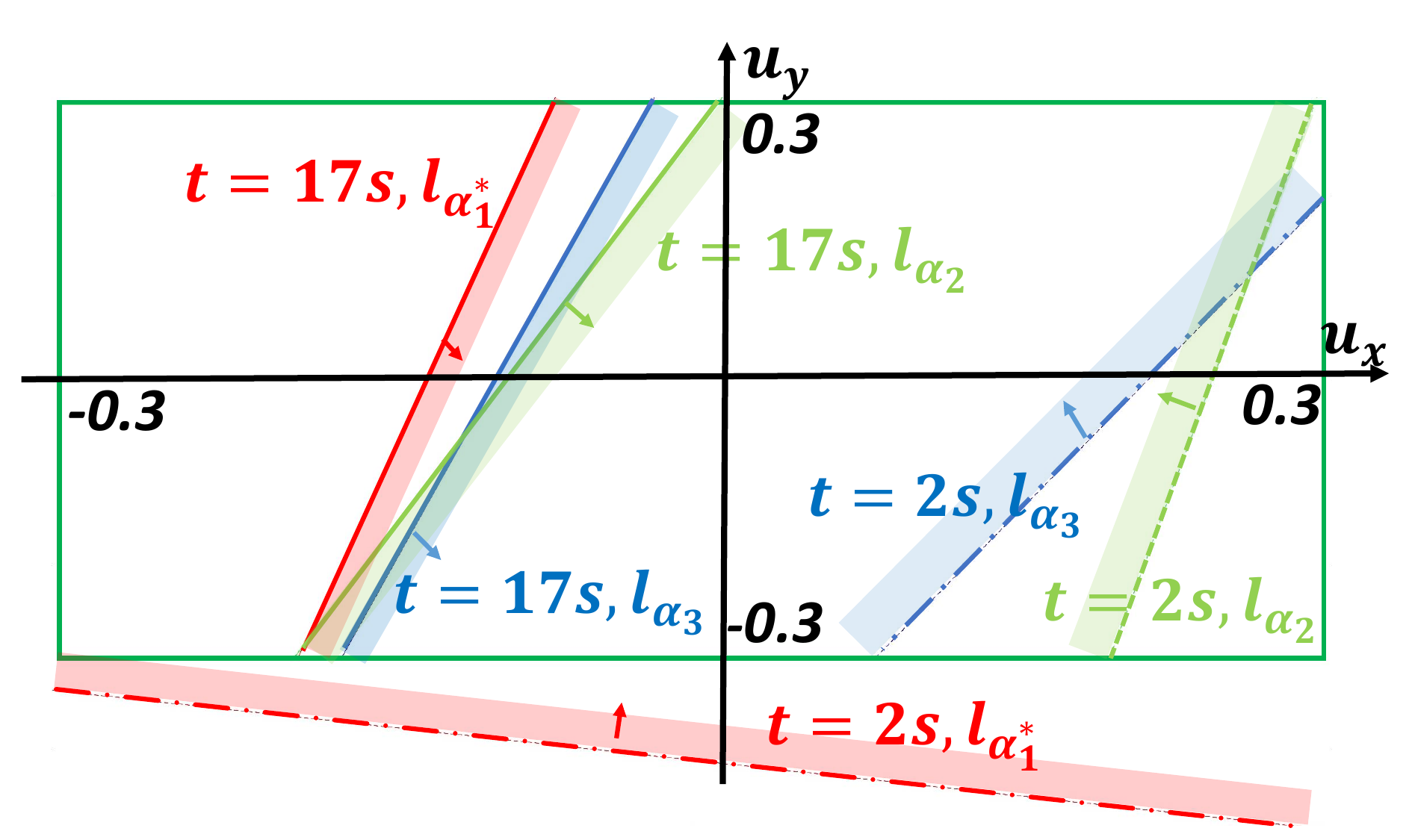}%
    \label{sim pd ACS}}
    \caption{The performance comparison between the optimized $\alpha^*_1$ and the predetermined $\alpha_2$, $\alpha_3$ associated QP solutions.}
    \label{sim comparision pd and optimized}
\end{figure}

 This subsection validates the effectiveness of our developed optimized ACS strategy \eqref{eq optimized ACS} clarified in Section \ref{sec Optimized ACS} based on a benchmark reach-avoid task. A mobile robot modelled as \eqref{eq control afine sys} is desired to move from an initial position $p_0$ to a desired position $p_d$ while avoiding one circle obstacle $\mathcal{O}$ (centered at $c=(1,1)$ and with radius $r = 1$) during the operation process, as illustrated in Fig.~\ref{sim pd traj}. The detailed simulation settings are referred to in TABLE \ref{tab parameter benchmark}.
Note that to avoid IL-CBFs and GD-CLFs' influence on the QP feasibility, 
this subsection uses a prior-known CBF to achieve collision avoidance, 
and an assumed nominal motion planning strategy to accomplish the reaching task with desired performance.
\begin{table}[ht]
  \begin{center}
    \caption{The parameter settings of the reach-avoid task.}
    \label{tab parameter benchmark}
    \begin{tabular}{c|c}
    \hline
      \multirow{1}{*}{\textbf{Initial values}} 
       &$p_0 = [-0.2,0.1]^{\top}$, $v_0 = [0,0]^{\top}$, $T = 10~\si{\Hz}$\\ \hline
       \multirow{1}{*}{\textbf{Target values}} 
       & $p_d = [2,1.5]^{\top}$,  $v_d = [0,0]^{\top}$\\ \hline
      \multirow{1}{*}{\textbf{CBF}} 
       & $h = (x-1)^2+(y-1)^2-1$ \\ \hline
    \multirow{1}{*}{\textbf{Nominal policy}} 
       & $u_{n} = -0.2(p-p_d)-0.9(v-v_d)$   \\ \hline
      \multirow{1}{*}{\textbf{QP and LP}} 
       & $\overline{u}_x$, $\overline{u}_y= 0.3$, $\alpha_1 (t_0) = [5,6]^{\top}$, $\overline{\alpha}_1 =7$\\ \hline
    \end{tabular}
  \end{center}
\end{table}

We formulate the QP optimization problem as 
\begin{subequations} \label{QP example PD}
    \begin{align}
    \min_{u}  &  \, \, \left\| u - u_{n} \right\|^2  \label{QP example PD 0}\\
\st    & \, \, -0.3 < u_x, u_y < 0.3  \label{QP example PD 1}\\
     & \, \, \Ddot{h} + \alpha^*_{1_{1}} \dot{h} + \alpha^*_{1_{2}} h \geq 0 \label{QP example PD 2},
\end{align}
\end{subequations}
 to solve the reach-avoid task mentioned above, where $\alpha^*_{1_{1}}$ and $\alpha^*_{1_{2}}$ are the optimized variables after solving the LP \eqref{eq optimized ACS} based on the known CBF $h$ presented in TABLE \ref{tab parameter benchmark}.
For comparison, prior-chosen constant vectors $\alpha_2 = [4,1]^{\top}$, $\alpha_3 = [4,2]^{\top}$ are picked to construct the constraint \eqref{QP example PD 2}. 
Note that the feasibility of the QP \eqref{QP example PD} is easily lost without choosing the suitable values of $\alpha$ required for the HO-CBF.  Here $\alpha_2$ and $\alpha_3$ are well-debugged parameters to ensure the QP feasibility.

As displayed in Fig. \ref{sim pd traj}, the nominal $u_{n}$ is an unsafe motion command given that the mobile robot driven by the $u_{n}$ crosses the obstacle $\mathcal{O}$. 
The minimally corrected $u_{n}$ by solving the QP \eqref{QP example PD} drives the mobile robot to safely reach the destination.
Furthermore, as shown in Fig. \ref{sim pd traj}, the trajectory of the optimized $\alpha_1^*$ case is closer to the desired trajectory (the cyan line) associated with $u_{n}$ as a consequence of the enlarged ACS.
The ACSs of the constraint \eqref{QP example PD 2} at $t = 2s$ and $t = 17s$ are displayed in Figure \ref{sim pd ACS}. It is shown that the $\alpha_1^*$'s associated ACS is larger than the related ones of $\alpha_2$ and $\alpha_3$. This validates the effectiveness of the LP optimization \eqref{eq optimized ACS}.

\subsection{Validation in Outdoor Scenario} \label{sec sim outdoor env}
\begin{figure}[h]
 \centering
    \subfloat[$t = 0.8s$.]{\includegraphics[width=1.8in]{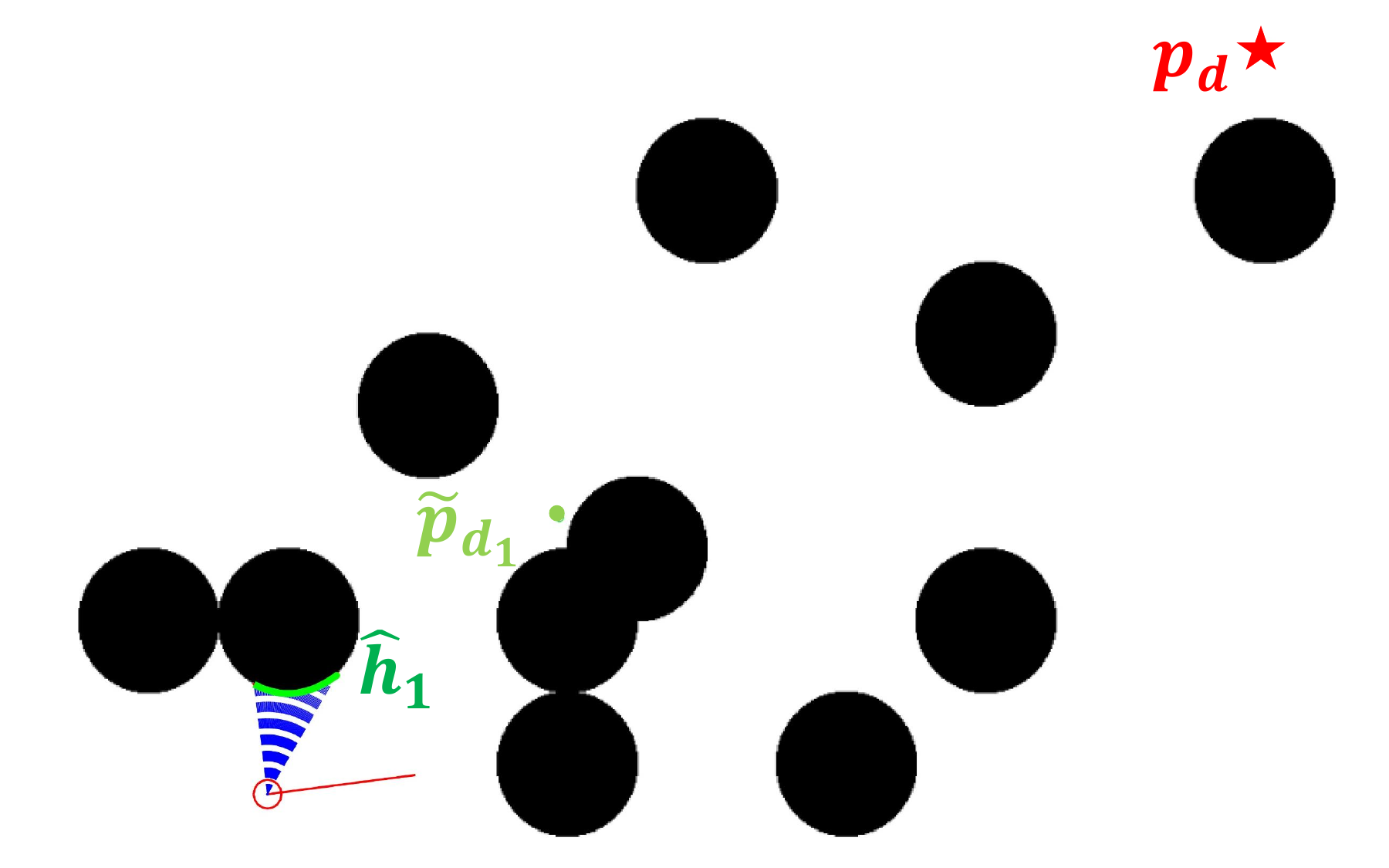}
    \label{fig t1  out}}
    \subfloat[$t = 2.3 s$]{\includegraphics[width=1.8in]{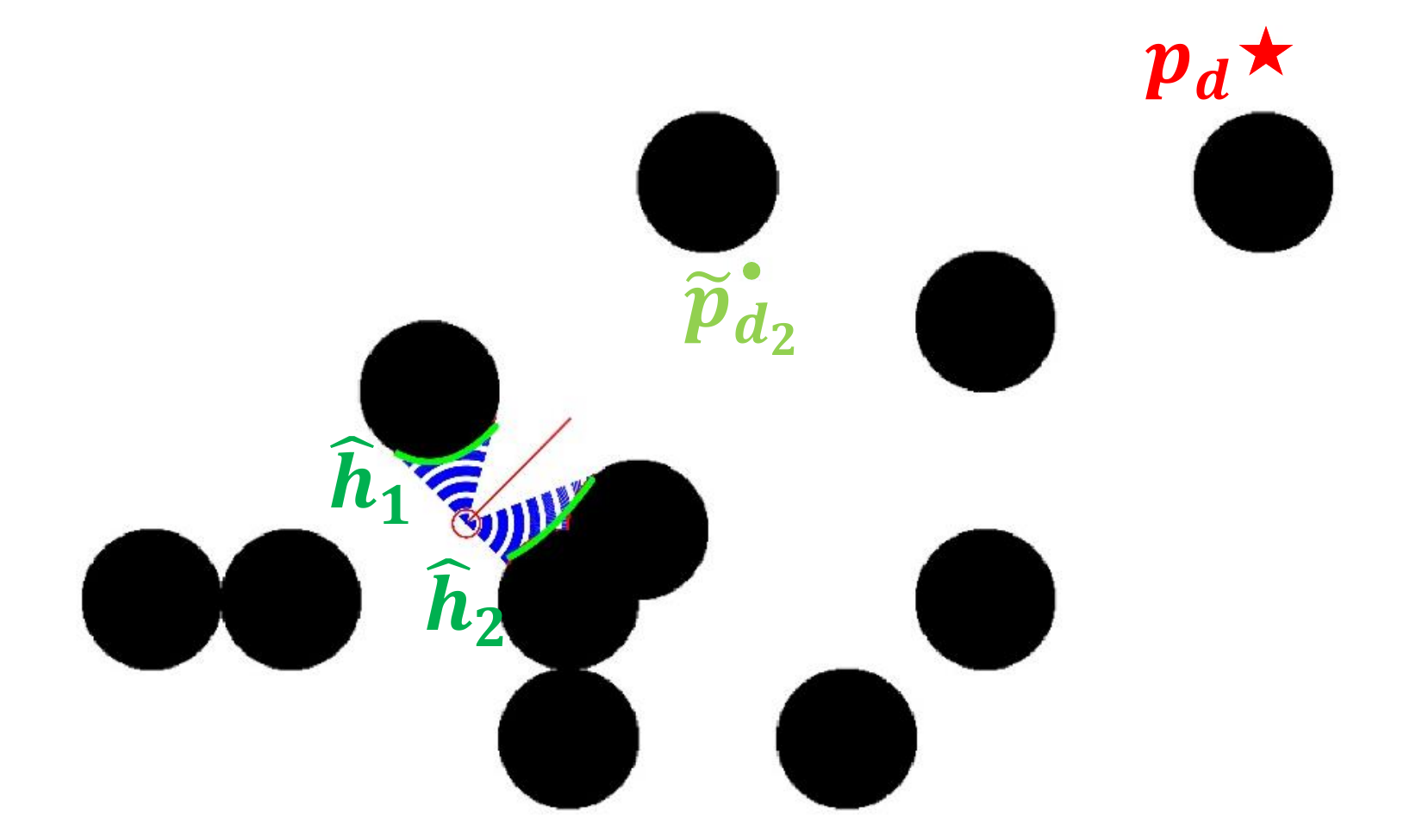}%
    \label{fig t2  out}}
    \\
    \subfloat[$t = 4.2 s$]{\includegraphics[width=1.8in]{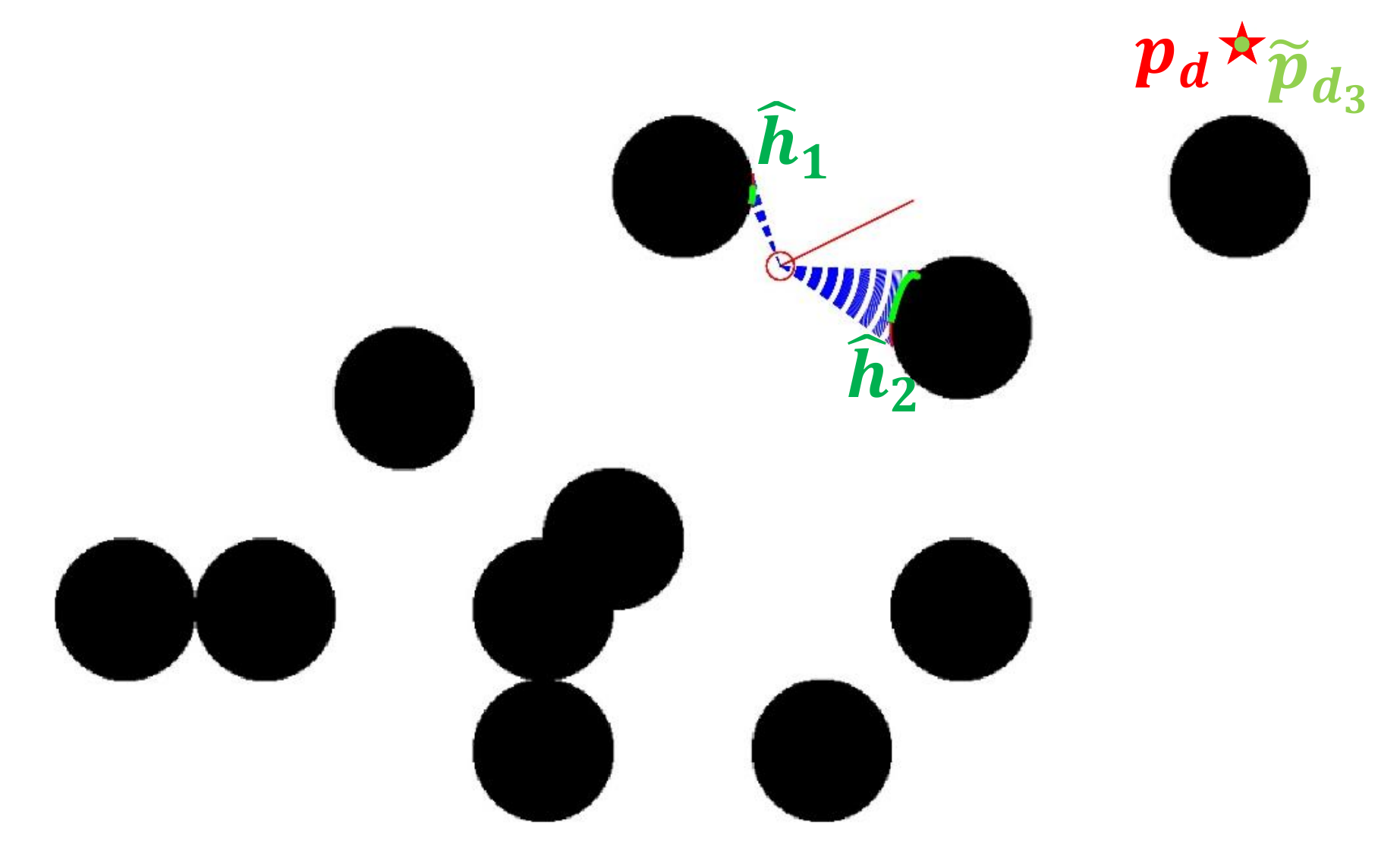}%
    \label{fig t5  out}}
     \subfloat[The whole trajectory of $p$.]{\includegraphics[width=1.8in]{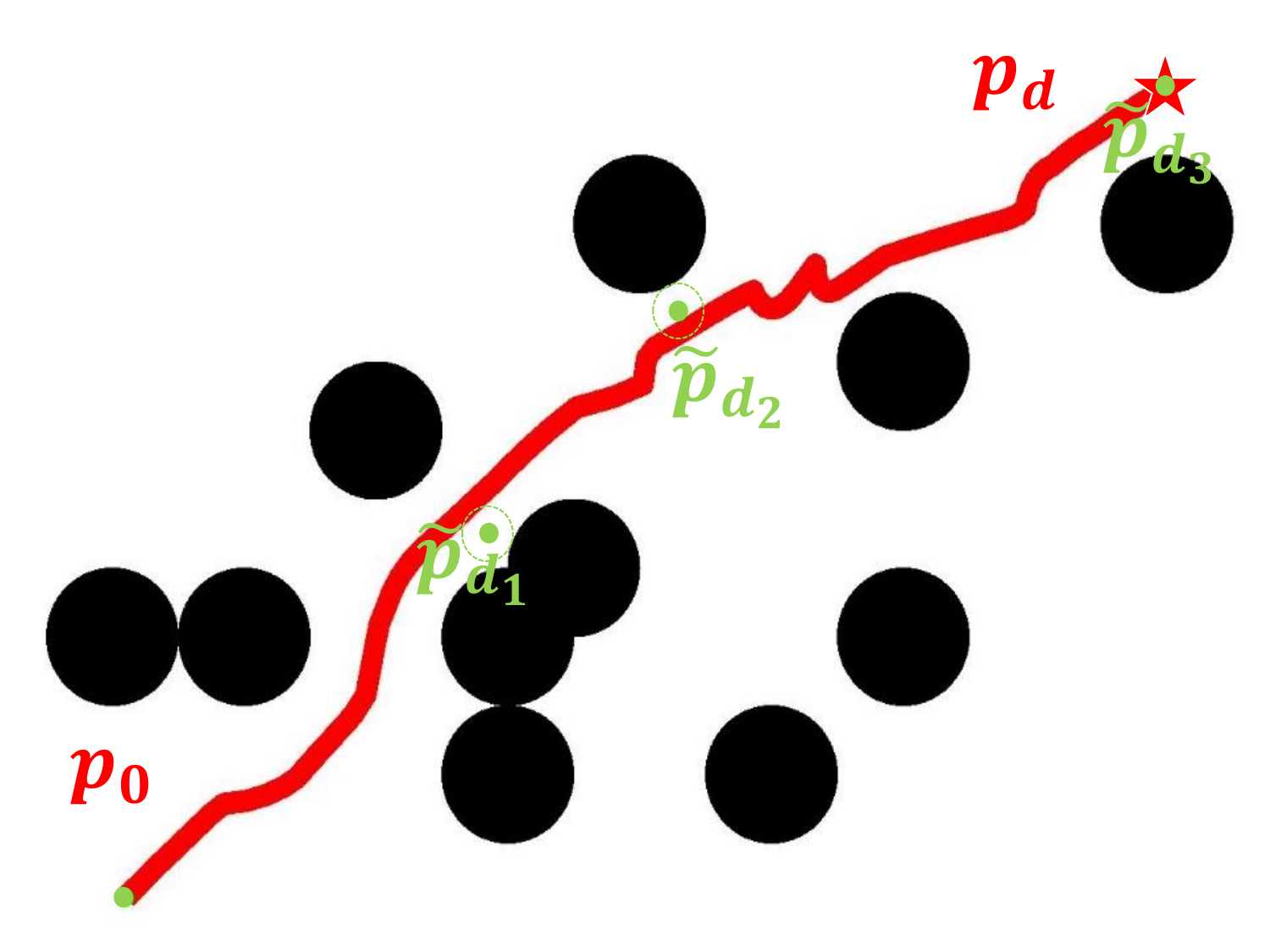}%
    \label{fig phasplot out}}
    \caption{
    The illustration of the robotic movement in the outdoor scenario. The black circles denote the prior-unknown obstacles; The dark green $\hat{h}_1$ and $\hat{h}_2$ denote the learned IL-CBFs; The light green $\tilde{p}_{d_{i}}$ represents the discovered subgoal;
    The light green $p_0$ and the red $p_d$ denote the initial and target positions, respectively.
    }
    \label{fig outdoor screenshots}
\end{figure}

This subsection validates the efficiency of our proposed SFMP strategy \eqref{eq QP} in an obstacle densely cluttered environment (see Fig. \ref{fig outdoor screenshots}). 
The numerical simulation is conducted on the basis of the Mobile Robotics Simulation Toolbox \cite{MRST} and the quadprog solver of the Optimization Toolbox \cite{coleman1999optimization}. The detailed parameter settings to solve the formulated QP \eqref{eq QP} and LP \eqref{eq optimized ACS} are presented in TABLE \ref{tab parameter outdoor navigation}.
\begin{table}[ht]
  \begin{center}
    \caption{The parameter settings of the outdoor scenario.}
    \label{tab parameter outdoor navigation}
    \begin{tabular}{c|c}
    \hline
      \multirow{1}{*}{\textbf{Initial values}} 
       &$p_0 = [2,4]^{\top}$, $v_0 = [1,1]^{\top}$, $T = 10~\si{\Hz}$\\ \hline
       \multirow{1}{*}{\textbf{Target values}} 
       & $p_d = [10,10]^{\top}$,  $v_d = [0,0]^{\top}$\\ \hline
      \multirow{1}{*}{\textbf{IL-CBF}} 
       & $\Phi = \left[1, x, x^2 \right]$, $S_{\theta} = [-\pi /2, \pi /2] $, $S_r = 0.5~m$ \\ \hline
    \multirow{1}{*}{\textbf{GD-CLF}} 
       & $P = \begin{bmatrix}
25 & 12.5 \\
12.5 & 25
\end{bmatrix}$, 
$Q = \begin{bmatrix}
50 & 25 \\
25 & 50
\end{bmatrix}$, \\
\multirow{1}{*}{} 
& $S_{\theta} = [-\pi, \pi]$, $S_r = 4~m$, $\bar{c}_2 = 1.5$ \\ \hline
      \multirow{1}{*}{\textbf{QP and LP}} 
       &$\overline{u}_x$, $\overline{u}_y= 20$, $\bar{c}_1 = 1$, $\alpha (t_0) = [5,6]^{\top}$, $\overline{\alpha} =6$\\ \hline
    \end{tabular}
  \end{center}
\end{table}

\begin{figure}[h]
 \centering
    \subfloat[The trajectory of input $u$.]{\includegraphics[width=1.8in]{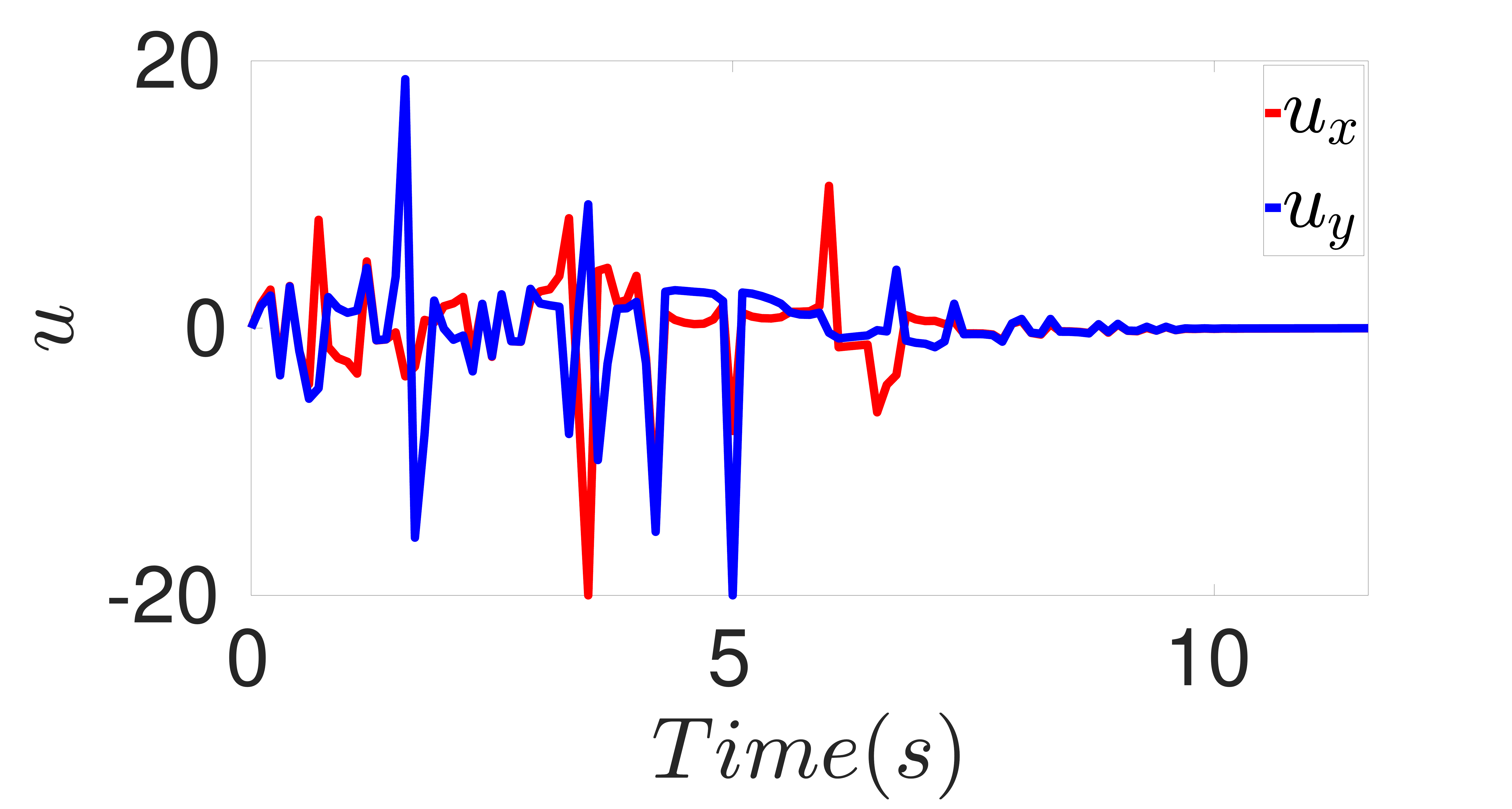}%
    \label{fig u out}}
    \subfloat[The trajectory of optimized $\alpha^*$.]{\includegraphics[width=1.8in]{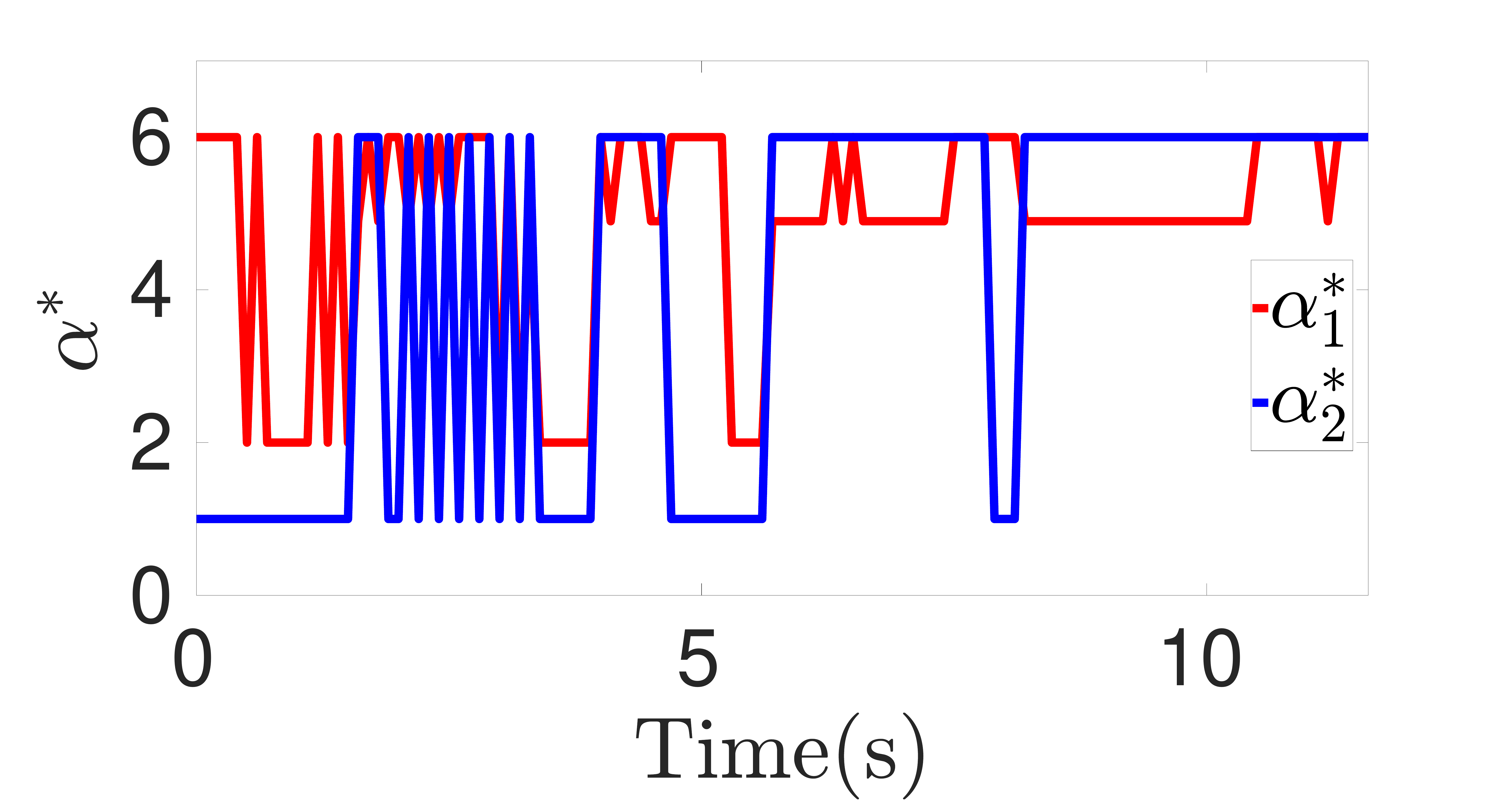}%
    \label{fig alpha out}}
    \caption{The trajectories of  $u$, and $\alpha^*$ for the outdoor scenario.}
    \label{fig outdoor case multiple states}
\end{figure}

It is shown in Fig. \ref{fig t1  out}-Fig. \ref{fig t5  out} that the mobile robot exploits sensed obstacle boundary data to learn the IL-CBFs $\hat{h}_1$, $\hat{h}_2$ using Algorithm \ref{LCBF learn algorithm}, and uses collision-free data to discover the subgoals $\tilde{p}_{d_{1}}$, $\tilde{p}_{d_{2}}$ via Algorithm \ref{GLCLF learn algorithm}. 
As displayed in Fig. \ref{fig phasplot out},
the mobile robot safely reaches the subgoals $\tilde{p}_{d_{1}}$, $\tilde{p}_{d_{2}}$ sequentially and finally reach the destination $p_d$ (same with $\tilde{p}_{d_{3}}$).
Thus, it is concluded that the learned IL-CBFs \eqref{eq IL-CBF} ensure collision avoidance with unforeseen obstacles,
and the constructed GD-CLFs \eqref{eq GLCLF} based on the discovered subgoals guarantee the task fulfillment. 
The evolution trajectories of the motion command $u$, and the optimized parameter $\alpha^*$ are displayed in Fig. \ref{fig u out} and Fig. \ref{fig alpha out}, respectively. 
The input saturation is satisfied, and the LP \eqref{eq optimized ACS} outputs the optimized $\alpha^*$ to ensure the feasibility of the QP \eqref{eq QP} during the whole operation process.
A supplemental video for the outdoor scenario is referred to in \url{https://youtu.be/FZsNc0UzEVs}.
\subsection{Validation in Indoor Scenario} \label{sec sim indoor env}
\begin{figure}[h]
    \centering
    \includegraphics[width=3.4in]{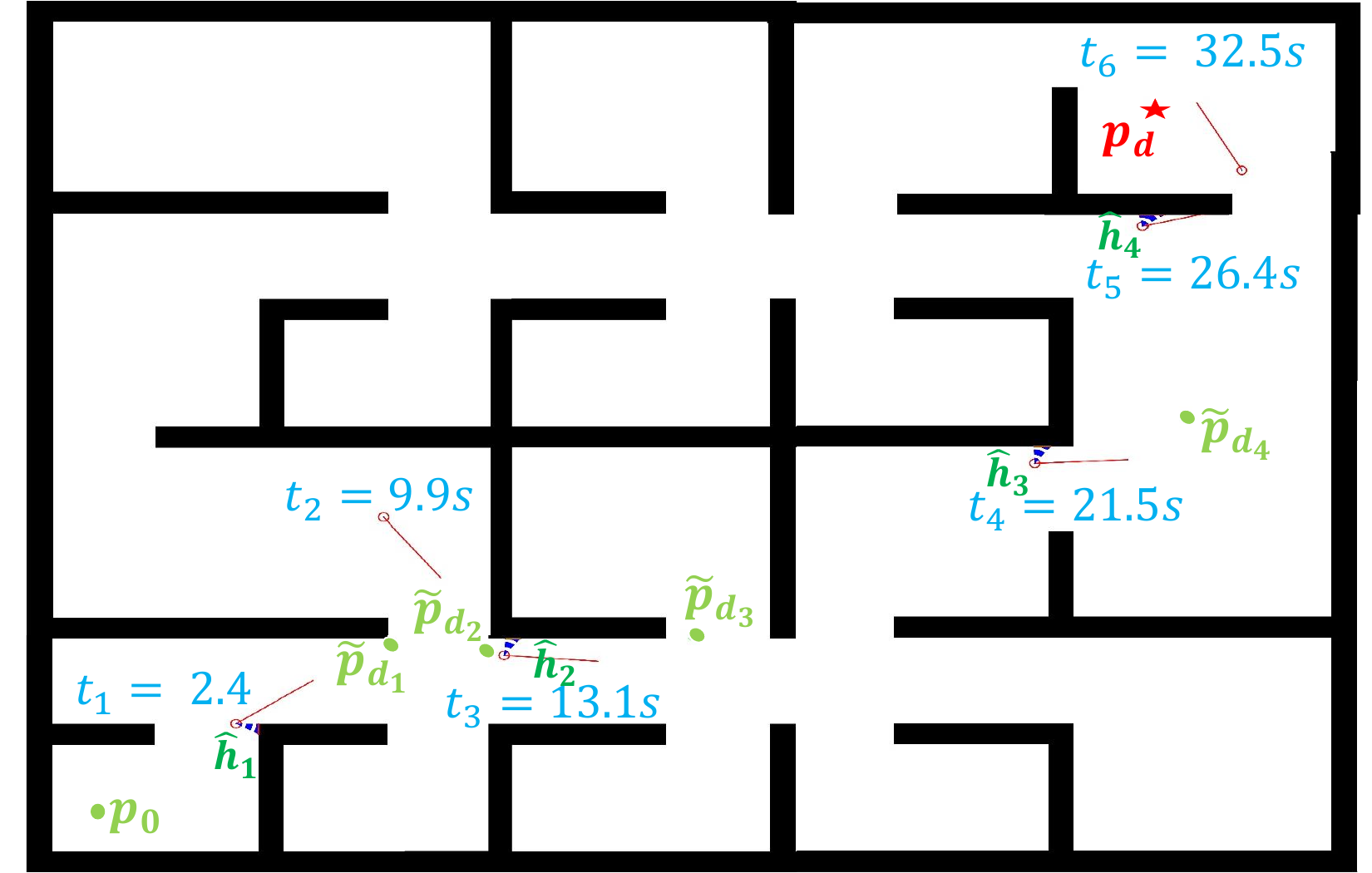}
    \caption{
    The illustration of the robotic movement in the indoor scenario.
    The thick black lines represent walls;
    The dark green $\hat{h}_i$ denotes the $i$-th learned IL-CBF;
    The light green $\tilde{p}_{d_{i}}$ represents the discovered $i$-th subgoal;
    The light green $p_0$ and the red  $p_d$ denote the initial and target positions, respectively.
    The pink line with a dot represents the position and the heading direction of the mobile robot at a specific time $t_i$ in light blue.
    }
    \label{fig whole in}
\end{figure}

This subsection further validates the effectiveness of our designed SFMP strategy \eqref{eq QP} in a maze simulation environment (see Fig. \ref{fig whole in}). 
It is worth mentioning that the application of common CBFs in a maze environment is seldom found in existing works.
This is because multiple typical CBFs are required to achieve collision avoidance in such a maze environment, and certain CBFs would unavoidably treat collision-free spaces as unsafe regions. 
In this case,  the mobile robot behaves conservatively and the QP might lose its feasibility.
In particular, for the maze environment displayed in Fig. \ref{fig whole in}, it is nontrivial to design barrier functions to separate safe and unsafe regions even though we have the full knowledge of the environment. However, our developed IL-CBFs can efficiently deal with this maze environment. The detailed parameters to realize the safe operation in the maze environment are provided in TABLE \ref{tab parameter indoor navigation}. Accompanying simulation videos are available at \url{https://youtu.be/FZsNc0UzEVs}.
\begin{table}[t]
  \begin{center}
    \caption{The parameter settings of the indoor scenario.}
    \label{tab parameter indoor navigation}
    \begin{tabular}{c|c}
    \hline
      \multirow{1}{*}{\textbf{Initial values}} 
       &$p_0 = [2,2]^{\top}$, $v_0 = [0,0]^{\top}$, $T = 10~\si{\Hz}$\\ \hline
       \multirow{1}{*}{\textbf{Target values}} 
       & $p_d = [22,18]^{\top}$,  $v_d = [0,0]^{\top}$\\ \hline
      \multirow{1}{*}{\textbf{IL-CBF}} 
       & $\Phi = \left[1, x, x^2 \right]$, $S_{\theta} = [-\pi /2, \pi /2]$, $S_r = 0.5~m $ \\ \hline
        \multirow{1}{*}{\textbf{GD-CLF}} 
       & $P = \begin{bmatrix}
25 & 12.5 \\
12.5 & 25
\end{bmatrix}$, 
$Q = \begin{bmatrix}
50 & 25 \\
25 & 50
\end{bmatrix}$, \\
\multirow{1}{*}{} 
& $S_{\theta} = [-\pi, \pi]$, $S_r = 4~m$, $\bar{c}_2 = 1.5$ \\ \hline
      \multirow{1}{*}{\textbf{QP and LP}} 
       &$\overline{u}_x$, $\overline{u}_y= 20$, $\bar{c}_1 = 1$, $\alpha (t_0) = [5,6]^{\top}$, $\overline{\alpha} =6$\\ \hline
    \end{tabular}
  \end{center}
\end{table}

\begin{figure}[h]
 \centering
     \subfloat[The trajectory of position $p$.]{\includegraphics[width=1.8in]{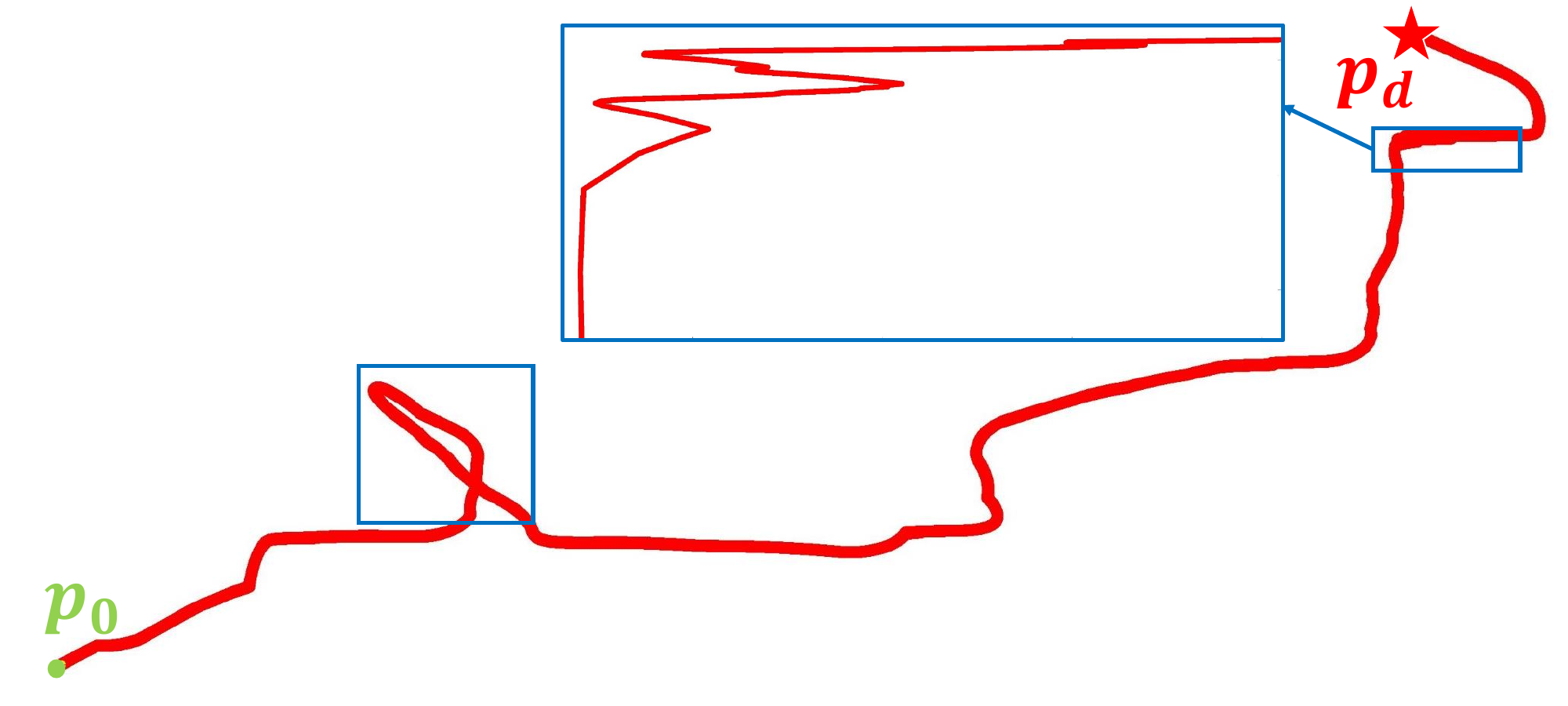}%
    \label{fig phasplot in}}
    \subfloat[The trajectory of velocity $v$.]{\includegraphics[width=1.8in]{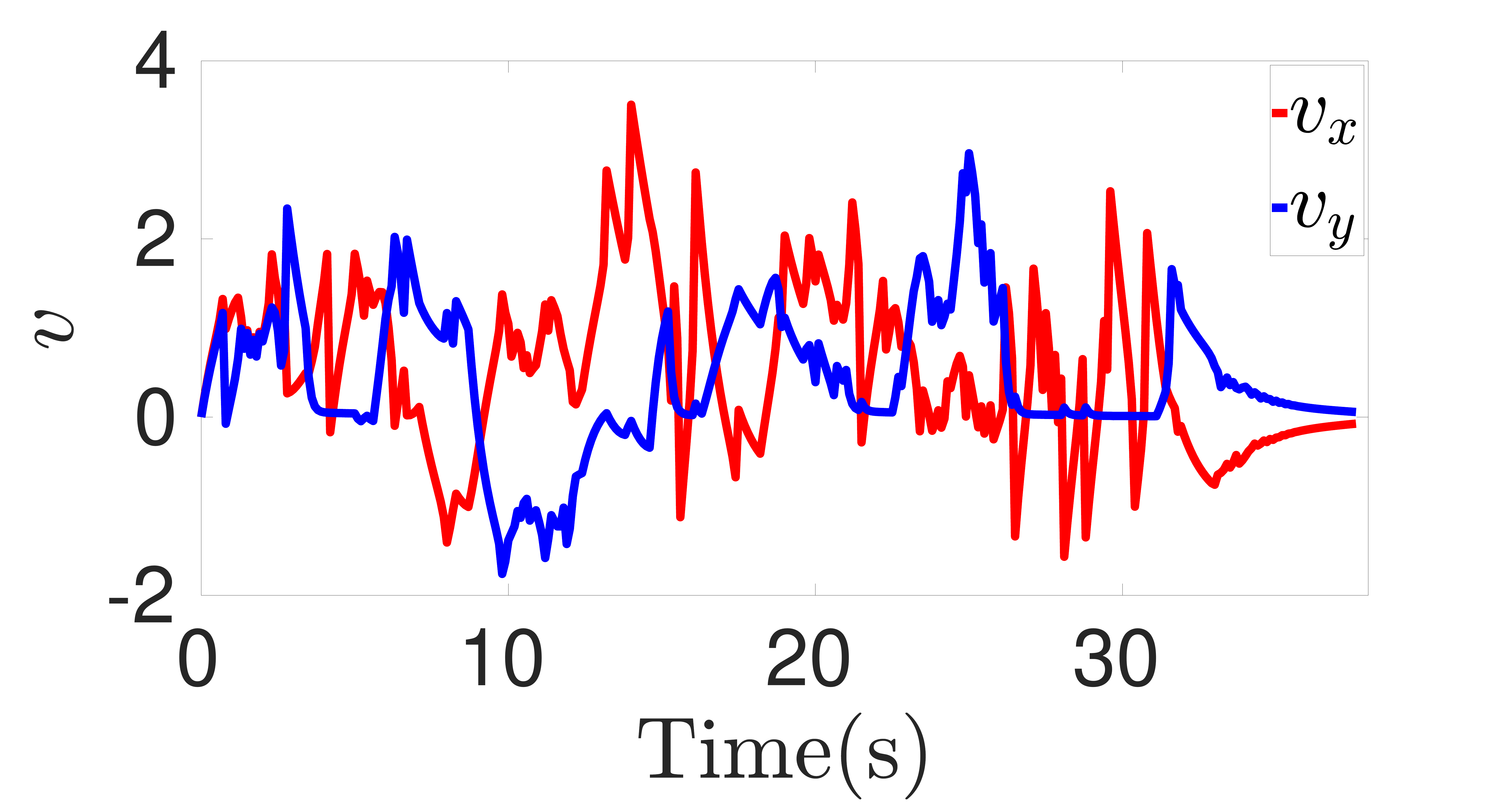}%
    \label{fig v in}}
    \\
    \subfloat[The trajectory of input $u$.]{\includegraphics[width=1.8in]{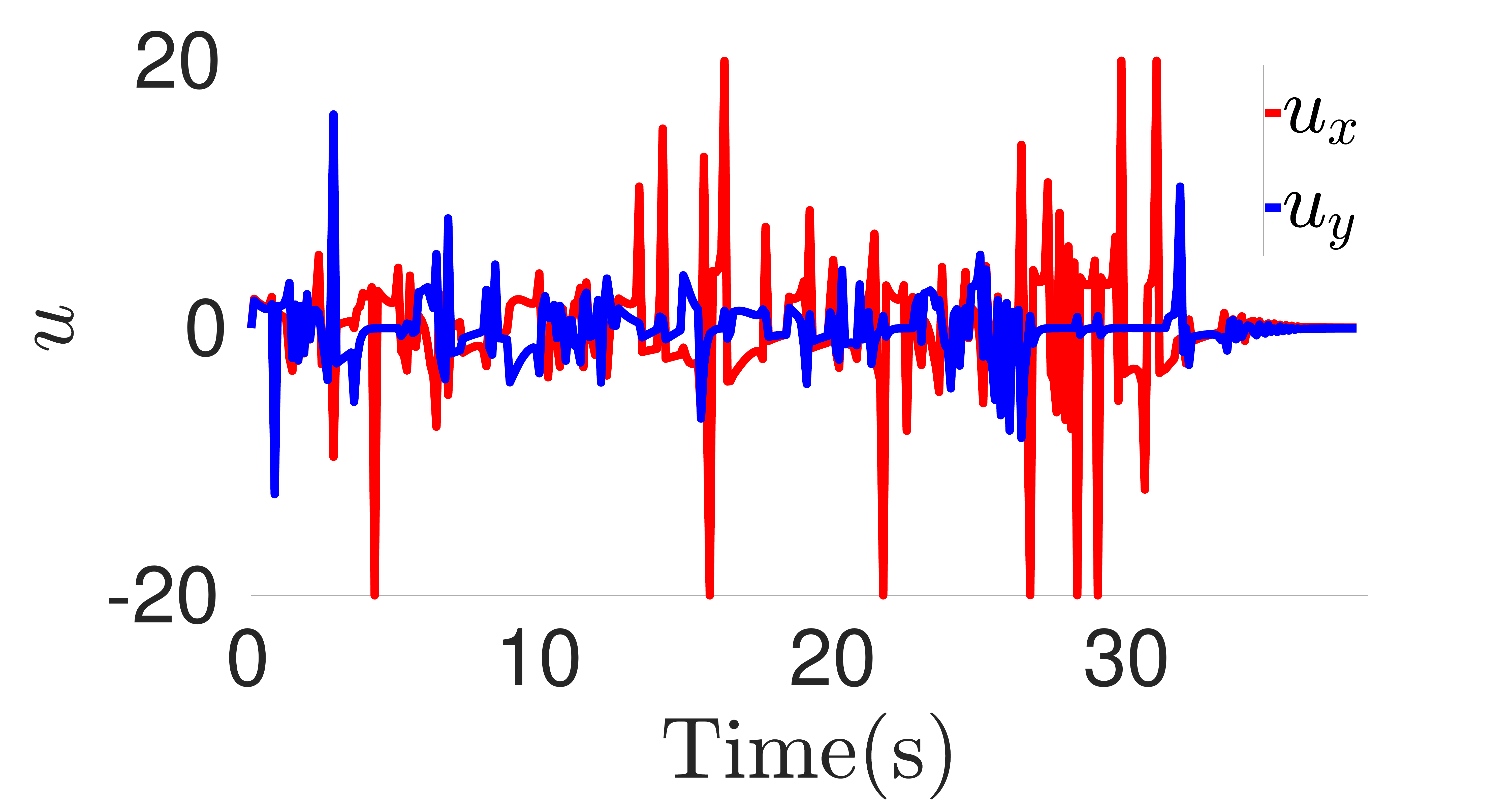}%
    \label{fig u in}}
    \subfloat[The trajectory of optimized $\alpha^*$.]{\includegraphics[width=1.8in]{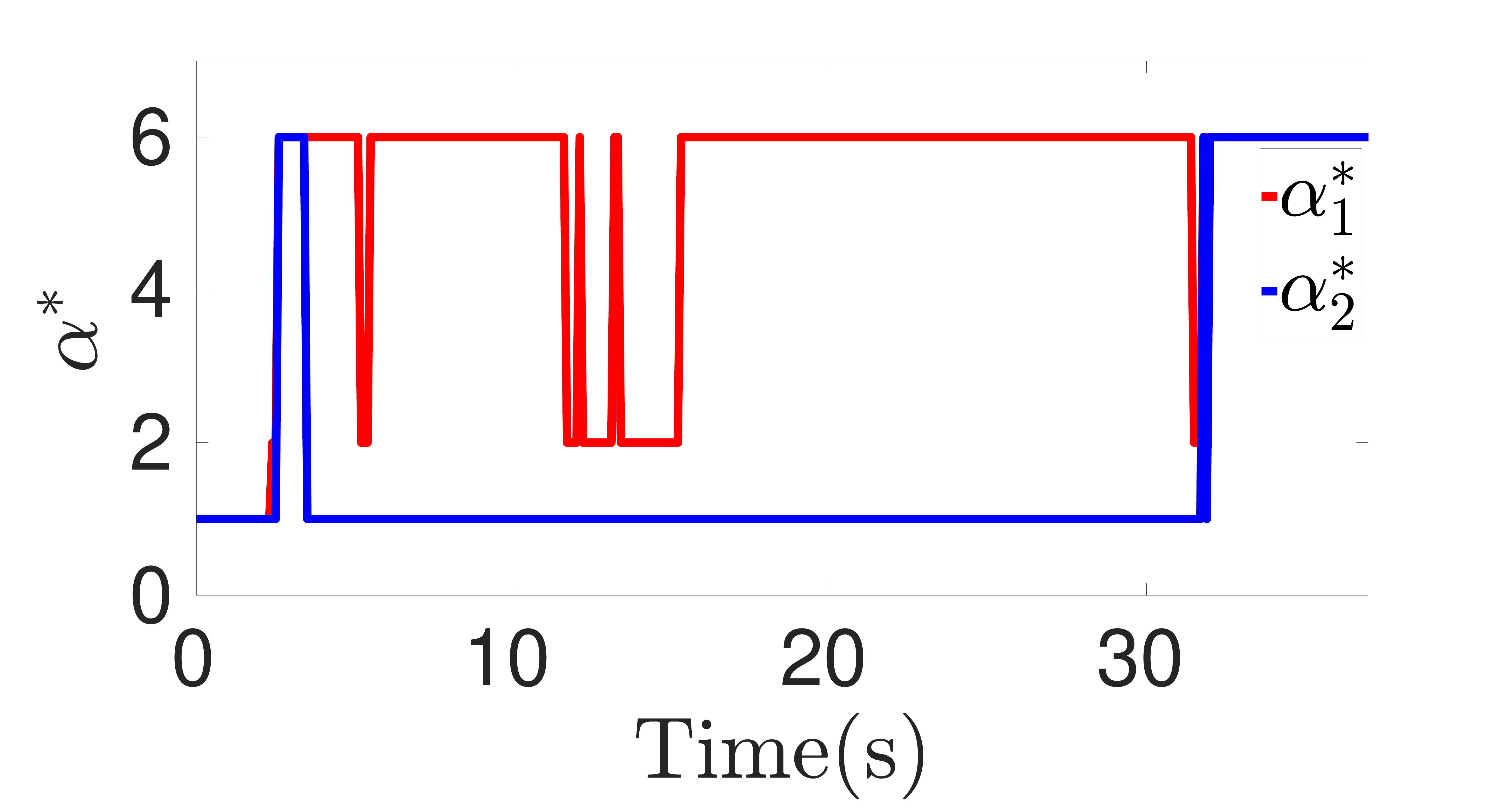}%
    \label{fig alpha in}}
    \caption{The trajectories of position $p$, velocity $v$, motion command $u$, and optimized $\alpha^*$ for the indoor scenario.}
    \label{fig indoor case multiple states}
\end{figure}

As displayed in  Fig. \ref{fig whole in}, the mobile robot operates safely in the maze environment and finally reaches the goal position $p_d$.
However, we observe inefficient operation (shown in the blue rectangle of Fig. \ref{fig phasplot in}) of the mobile robot in this unforeseen maze environment.
This is due to the simple heuristic (the shortest distance rule in particular) used in Algorithm \ref{GLCLF learn algorithm}.
This problem could be avoided by changing the sensor range in an adaptive way. We deliberately present this incomplete case to show the potential drawback of our method.
The trajectories of the mobile robot's velocity $v$, motion command $u$ and optimized $\alpha^*$ are displayed in Fig. \ref{fig v in}, Fig. \ref{fig u in}, and Fig. \ref{fig alpha in}, respectively.
The input saturation is always satisfied and $\alpha^*$ updates to ensure the QP feasibility.

\subsection{Validation in High-fidelity Simulator} \label{sec ros simu}
We further evaluate the performance of our proposed SFMP strategy \eqref{eq QP} in different scenarios based on Gazebo \cite{koenig2004design} and robot operating system (ROS) \cite{quigley2009ros}. 
The simulations are conducted on the Ubuntu 18.04 computer with 16 GB RAM and 2.6-GHz Intel Core i7-9750H CPU.
The adopted Mecanum wheel cart is equipped with a LiDAR sensor ( $S_{\theta} = [-\pi /4, \pi /4]$, $S_r = 4~m$) to perceive the surrounding environment.
To account for the robot volume's influence on safety,  the original detected obstacle boundary samples are projected backwards along the LiDAR laser line by a distance $d = 0.5~m$.

\subsubsection{Basic Demos}
\begin{figure}[!t]
    \centering
    \includegraphics[width=3.4in]{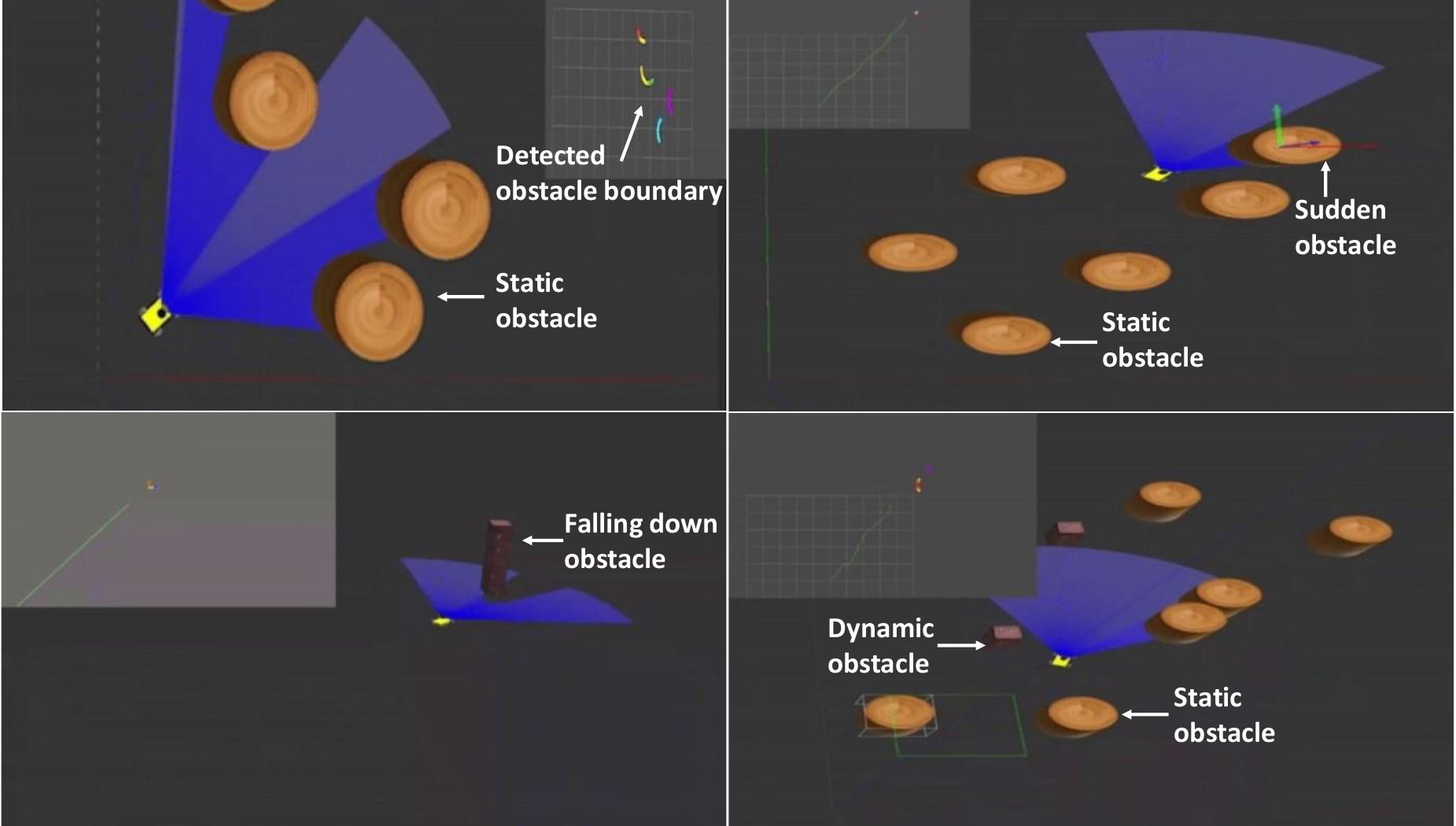}
    \caption{The validation of the SFMP strategy in different simulated environments.
    The yellow cart uses a sensor with a limited detection range shown in blue to perceive the environment, wherein the cylinders denote the static obstacles, and the cuboids present the dynamic obstacles.}
    \label{fig ros different case}
\end{figure}

The effectiveness of our developed SFMP strategy \eqref{eq QP} for the safe operation in unknown environments is validated via the following purposely designed cases: 
(a) static obstacles (the top-left Fig.~\ref{fig ros different case}); 
(b) static plus suddenly added static obstacles (the top-right Fig.~\ref{fig ros different case});
(c) falling down dynamic obstacles (the bottom-left Fig.~\ref{fig ros different case}); 
and (d) static plus dynamic obstacles (the bottom-right Fig.~\ref{fig ros different case}). 
Note that we simply extend the learned IL-CBFs \eqref{eq IL-CBF} to avoid collision with dynamic obstacles here. Although without rigorous analysis, we found that the learned IL-CBFs could also address slowly moving obstacles. 
This is because we use instantaneous sensory data (reflecting the environmental changes timely) for collision avoidance.
The results shown in Fig.~\ref{fig ros different case} and the associated video at \url{https://youtu.be/bpWW9R_MYpc} validate that our proposed SFMP strategy \eqref{eq QP} would drive the cart to survive in these four first-entry environments populated with static and dynamic obstacles.
\subsubsection{Large Environment}
\begin{figure}[!t]
    \centering
    \includegraphics[width=3.4in]{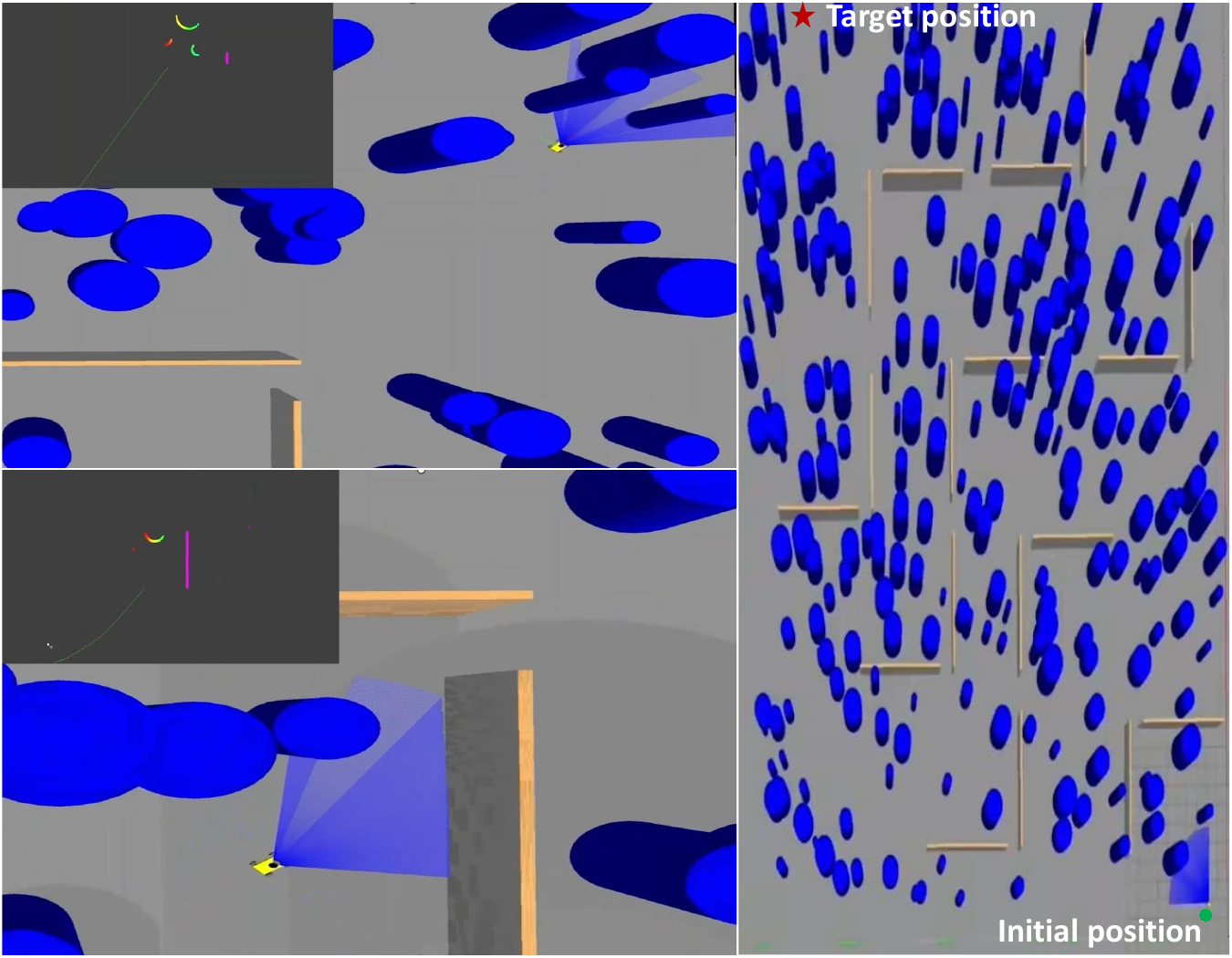}
    \caption{The validation of the proposed SFMP strategy in a large environment. The map size is $50~ \si{\metre} \times 50~ \si{\metre}$.
    Left: the screenshots of the mobile robot and the environment at specific time instants;
    Right: the top view of the task and the environment;
    The blue cylinders and brown walls denote obstacles;
    The green point and the red pentagram denote the initial and target positions, respectively.}
    \label{fig ros large env}
\end{figure}

We further examine the performance of the SFMP strategy \eqref{eq QP} in a large environment (a mix of outdoor and indoor scenarios), see Fig.~\ref{fig ros large env} and the associated video in \url{https://youtu.be/hDdyKatrkCA}. 
We randomly sample 10 initial positions in the circle with center$c_0 =(0,0)$ and radius $r_0 = 5~m$, and 10 goal positions in the circle with center $c_t =(45,45)$ and radius $r_t = 5~m$.
The proposed SFMP strategy succeeds 9 times out of 10.
The success to survive in this complex environment without collisions and finally reaching the target position proves the practicability of our method regarding the robustness towards varying tasks (represented as different initial and target positions).
Note that we purposely use the bottom-left initial position and the top-right target position in Fig.~\ref{fig ros different case} to simulate a long-horizon task and also encourage the mobile robot to meet more obstacles.
\subsubsection{Comparative Evaluation}
\begin{figure}[!t]
    \centering
    \includegraphics[width=3.4in]{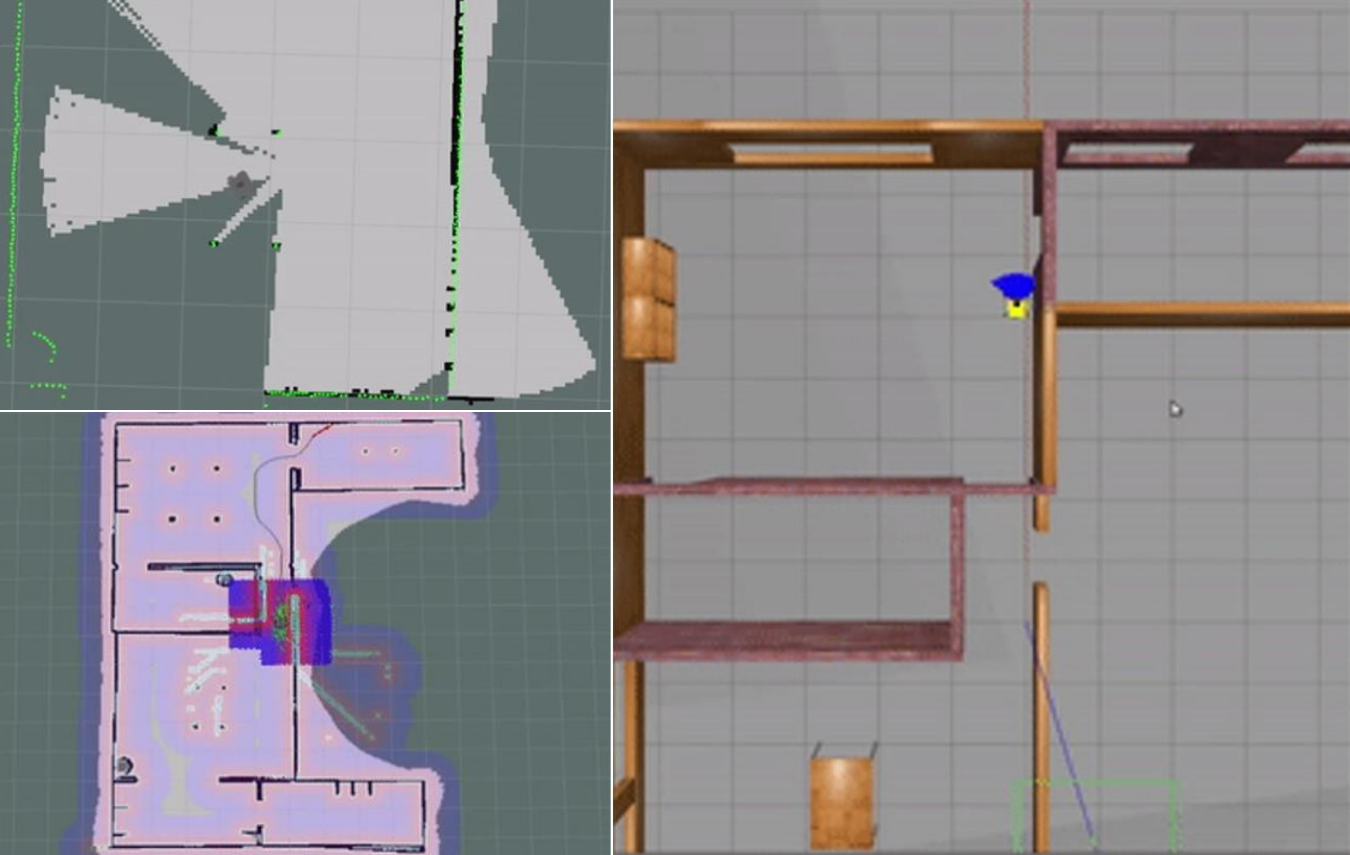}
    \caption{The comparative evaluations conducted in a room in the size $12~\si{\metre} \times 16~\si{\metre}$.
    The left figures are snapshots related to the baseline RRT method.
    Top left: the time-consuming mapping process;
    Bottom left: the movement of the cart on a pre-built map.
    The right figure is a snapshot of the movement of the cart driven by our method.}
    \label{fig ros comparision}
\end{figure}

This part focuses on the safe execution task in a room to show the superiority of our SFMP strategy \eqref{eq QP} over the baseline rapidly exploring random tree (RRT) method \cite{RRTApp} (a common sampling based motion planning strategy) in terms of time.  A supplementary video is referred to in \url{https://youtu.be/lelW7C_mfsE}.
For the scenario displayed in Fig.~\ref{fig ros comparision}, the common approach used in \cite{RRTApp} would firstly build a perfect map using the SLAM technique (1025 seconds in the top-left Fig.~\ref{fig ros comparision}) and then RRT generates a collision-free path followed by a PD controller (70 seconds in the bottom-left Fig.~\ref{fig ros comparision}). In summary, the common approach would consume 1025 seconds in total to generate a safe solution in this room scenario.
Our developed approach (the right Fig.~\ref{fig ros comparision}) perceives the local environment and outputs the SFMP strategy \eqref{eq QP} that drives the cart from the initial position to the target position. The utilized total time is 114 seconds.
Regarding the first-entry environment, especially with no need to build a perfect global map, our SFMP strategy \eqref{eq QP} enjoys an obvious advantage regarding time.  

\section{Conclusion} \label{sec conclusion}
This work presents a safe feedback motion planning strategy that fulfils the nontrivial safe operation in prior-unknown environments.
Our developed instantaneous local control barrier functions are united with goal-driven control Lyapunov functions in a quadratic programming optimization framework to generate safe feedback motion planning strategies.
The formulated linear programming optimization enhances the quadratic programming solution feasibility by enlarging the admissible control spaces of instantaneous local control barrier functions.
Multiple conducted numerical validations fully prove the effectiveness of our proposed safe feedback motion planning strategy.
The future work aims to extend our developed instantaneous local control barrier functions to realize collision avoidance with dynamic obstacles within consideration of obstacles' velocity and size.
Besides, fully exploiting the maneuverability of the mobile robot to reach the target position in a time-optimal way is well worth investigating.
\vspace{1cm}

\noindent \textbf{Author Contributions} Cong Li contributed to the first draft of the manuscript and the development of the theoretical contributions.
Zengjie Zhang contributed to the analysis and interpretation of the
results. 
Nesrin Ahmed contributed to implementing the numerical validations.
Qingchen Liu, Fangzhou Liu and Martin Buss supervised the study design,
and reviewed, edited, and prepared the final version of the paper.

\vspace{0.2cm}
\noindent \textbf{Funding} This research is supported by the Science and Technology Major Project of Anhui Province (202203a06020011)

\vspace{0.2cm}
\noindent \textbf{Data Availability} Data and materials used are available by contacting the corresponding author.

\vspace{0.2cm}
\noindent \textbf{Code Availability} The complete simulation data is available by contacting the corresponding author.

\section*{Declarations}
\noindent \textbf{Ethics Approval}  Not applicable.

\vspace{0.2cm}
\noindent \textbf{Consent to Participate} Not applicable.

\vspace{0.2cm}
\noindent \textbf{Consent for Publication} All authors have approved and consented to publish the manuscript.

\vspace{0.2cm}
\noindent \textbf{Conflict of Interests}
The authors have no relevant financial or nonfinancial interests to disclose.

\bibliographystyle{IEEEtran}
\bibliography{bibtex/bib/IEEEexample}

\noindent \textbf{Cong Li} received the Ph.D. degree in learning-based control and robotics with the Chair of Automatic Control Engineering, Technical University of Munich, Munich, Germany. He is also a Research Associate with the Chair of Automatic Control Engineering, Technical University of Munich. His research interests include reinforcement learning, optimal control, robust control, constraint optimization, and robotics.
\vspace{0.2cm}

\noindent \textbf{Zengjie Zhang}  received his Bachelor and Master
degrees from Harbin Institute of Technology, China, in 2013 and 2015 respectively. 
He received the Doktor-Ingenieur degree in electrical engineering from the Technical University of Munich, Germany, in 2020. 
He was a Research Associate at the Chair of Automatic Control Engineering of the Technical University of Munich, Germany while pursuing the doctoral degree. 
Now he is within the Department of Electrical Engineering, Eindhoven University of Technology, The Netherlands. 
His research interests include sliding mode control, fault detection and isolation and human-robot collaboration.
 \vspace{0.2cm}
 
\noindent \textbf{Nesrin Ahmed} received his Bachelor and Master
degrees from Technical University of Munich, Germany, in 2017 and 2020 respectively. Her interests include optimal control and collision avoidance.
 \vspace{0.2cm}
 
 \noindent \textbf{Qingchen Liu}
  received the Ph.D degree in system and control from the Australian National University, Canberra, Australia, in 2018. From 2018 to 2019, he worked as a Postdoc Research Fellow in the same group. From 2019-2021, he was an EuroTech Research Fellow at the Chair of Information-Oriented Control, Technical University of Munich, Munich, Germany. He is now a Research Associate Professor at the Department of Automation, University of Science and Technology of China. His research interest includes networked systems and multi-robotics systems.
\vspace{0.2cm}

 \noindent \textbf{Fangzhou Liu}
 received the M.Sc. degree in control theory and engineering from Harbin Institute of Technology, Harbin, China,
in 2014, and the Doktor-Ingenieur degree in electrical engineering from the Technical University
of Munich, Germany, in 2019. 
He was a Lecturer and a Research Fellow with the Chair of Automatic Control Engineering, Technical University of Munich, Munich, Germany. He is now a Full Professor at the School of Astronautics, Harbin Institue of Technology, Harbin, China. He has received the Dimitris N. Chorafas Prize and the Promotionspreis der Fakultät für Elektrotechnik und Informationstechnik. His current research interests include networked control systems; modeling, analysis, and control on social networks; reinforcement learning and their applications.
\vspace{0.2cm}

 \noindent \textbf{Martin Buss}
received the Diploma
Engineering degree in electrical engineering from
the Technische Universität Darmstadt, Darmstadt,
Germany, in 1990, and the Doctor of Engineering
degree in electrical engineering from The University
of Tokyo, Tokyo, Japan, in 1994.

In 1988, he was a Research Student for one year
with the Science University of Tokyo. From 1994 to
1995, he was a Post-Doctoral Researcher with the
Department of Systems Engineering, The Australian
National University, Canberra, ACT, Australia. From
1995 to 2000, he was a Senior Research Assistant and a Lecturer with the
Chair of Automatic Control Engineering, Department of Electrical Engineering and Information Technology, Technical University of Munich, Munich,
Germany. From 2000 to 2003, he was a Full Professor, the Head of the
Control Systems Group, and the Deputy Director of the Faculty IV, Electrical
Engineering and Computer Science, Institute of Energy and Automation
Technology, Technical University of Berlin, Berlin, Germany. Since 2003,
he has been a Full Professor (Chair) with the Chair of Automatic Control
Engineering, Faculty of Electrical Engineering and Information Technology,
Technical University of Munich, where he has been with the Medical Faculty
since 2008. His research interests include automatic control, mechatronics,
multimodal human system interfaces, optimization, nonlinear, and hybrid
discrete-continuous systems.
\end{document}